\documentclass[12pt]{article}
\evensidemargin=\oddsidemargin
\usepackage{color}
\usepackage{cite}
\usepackage[parfill]{parskip}

\definecolor{Red}           {cmyk}{0,1,1,0}
\definecolor{Black}         {cmyk}{0,0,0,1}

\newcounter{bla}
\newenvironment{refnummer}{%
\list{[\arabic{bla}]}%
{\usecounter{bla}%
 \setlength{\itemindent}{0pt}%
 \setlength{\topsep}{0pt}%
 \setlength{\itemsep}{0pt}%
 \setlength{\labelsep}{2pt}%
 \setlength{\listparindent}{0pt}%
 \settowidth{\labelwidth}{[9pt]}%
 \setlength{\leftmargin}{\labelwidth}%
 \addtolength{\leftmargin}{\labelsep}%
 \setlength{\rightmargin}{0pt}}}
 {\endlist}

\newcommand{\imag}{\Im {\rm m}}
\newcommand{\real}{\Re {\rm e}}

\newcommand{\ra}{\rightarrow}
\newcommand{\bea}{\begin{eqnarray}}
\newcommand{\eea}{\end{eqnarray}}

\newcommand{\lsim}{\raisebox{-0.13cm}{~\shortstack{$<$ \\[-0.07cm] $\sim$}}~}
\newcommand{\gsim}{\raisebox{-0.13cm}{~\shortstack{$>$ \\[-0.07cm] $\sim$}}~}
\newcommand{\bra}[1]{\langle #1|}
\newcommand{\ket}[1]{|#1\rangle}

\newcommand{\bc}{\begin{center}}
\newcommand{\ec}{\end{center}}

\def\beq{\begin{equation}}
\def\eeq{\end{equation}}
\def\beas{\begin{eqnarray*}}
\def\eeas{\end{eqnarray*}}
\def\ba{\begin{array}}
\def\ea{\end{array}}

\def\code{{\tt SUSY\_FLAVOR}}

\def\webpage{{\tt http://www.fuw.edu.pl/susy\_flavor}}

\begin{document}

\thispagestyle{empty}

\begin{flushright}
{\tt IPPP-10-16}\\ 
{\tt DCPT-10-32}\\ 
{\tt arXiv:1003.4260} \\
\today
\end{flushright}

\bigskip

\begin{center}
{\Large \bf {\color{Red}\code~}{\color{Black}: a computational tool for
FCNC and \\[3mm] CP-violating processes in the MSSM}}
\\[1.7cm]

{\large
J.~Rosiek$^{a,b}$,
P.~H.~Chankowski$^a$, 
A.~Dedes$^{c,b}$, 
S.~J\"ager$^d$ 
and P.~Tanedo$^{e,b}$}\\[0.5cm]

  {\em $^a$Institute of Theoretical Physics, University of Warsaw,
    Ho\.za 69, 00-681 Warsaw, Poland}\\[0.2cm]
  {\em $^b$Institute for Particle Physics Phenomenology, University of
    Durham, DH1 3LE, UK}\\[0.2cm]
  {\em $^c$Division of Theoretical Physics, University of Ioannina, GR
    45110, Greece}\\[0.2cm]
  {\em $^d$ Department of Physics and Astronomy, University of Sussex,
     Brighton BN1 9QH, UK}\\[0.2cm]
  {\em $^e$Institute for High Energy Phenomenology, Newman Laboratory
    of Elementary Particle Physics, Cornell University, Ithaca, NY
    14853, USA}

\end{center}

\vspace*{0.8cm}

\centerline{\bf ABSTRACT} 

\medskip\noindent We present \code{} --- a Fortran 77 program that
calculates important leptonic and semi-leptonic low-energy observables
in the general $R$-parity conserving MSSM.  For a set of input MSSM
parameters, the code gives predictions for the $\bar K^0 K^0$, $\bar D
D$, $\bar B_d B_d$ and $\bar B_s B_s$ mixing parameters; $B\ra X_s
\gamma$, $B_{s,d}\ra l^+ l^-$, $K^0_L\ra \pi^0 \bar\nu\nu$ and $K^+\ra
\pi^+ \bar\nu\nu$ decay branching ratios; and the electric dipole
moments of the leptons and the neutron.  All these quantities are
calculated at one-loop level (with some higher-order QCD corrections
included) in the exact sfermion mass eigenbasis, without resorting to
mass insertion approximations.
The program can be obtained from \webpage.

\newpage
\setcounter{page}{1}

\section{Introduction}
\label{sec:intro}

Flavor changing neutral currents (FCNCs) in the Minimal Supersymmetric
Standard Model (MSSM)~\cite{reviewsMSSM} originate from the fact that
one cannot, in general, simultaneously diagonalize the mass matrices
of fermions and their supersymmetric partners.
The misalignment between these mass matrices leads to FCNCs at tree
level.  Moreover a large number of the MSSM parameters which can take
complex values is a potential source of CP violation. Thus
supersymmetric contributions to amplitudes of processes violating
flavor and to quantities measuring CP violation, like Electric Dipole
Moments (EDMs), can exceed by orders of magnitudes the ones of the SM
particles.
Such large effects are ruled out by experimental measurements which
generally agree with the SM predictions and thus provide strong bounds
on the amount of flavor and CP violation in the MSSM.  For instance,
measurements of the kaon system properties prohibit FCNC couplings
between the first and second generation of down-type squarks larger
than $10^{-3}$.  These strong limits are often called the ``SUSY
flavor problem''.
However, there are also areas where current experiments still leave
room for large SUSY contributions.  For example, constraints from
$B$-meson experiments allow an $\mathcal{O}(1)$ mixing between the
second and third generation down-type squarks.  Such a large mixing
could produce FCNC effects that could be observed in the future at
$B$-factories and/or hadron colliders, like the Tevatron and
LHC~\cite{Buras}.

Even assuming the so-called Minimal Flavor Violation (MFV) scenario in
which {\it all} FCNC effects originate from the superpotential Yukawa
couplings, the flavor conserving soft SUSY breaking parameters can
still contain complex phases that cannot be absorbed by a redefinition
of fields and can, for example, give large contributions to the
electron and neutron EDMs.

As the accuracy of rare decay experiments improves, it is important to
have a universal computational tool which would help compare new data
with the predictions of the MSSM.  Constructing such a tool is a
non-trivial task because finding SUSY contribution to each rare decay
requires tedious calculations, especially when one wishes to have
fully general formulae that do not rely on the restrictive assumptions
of the MFV scenario.  Numerous analyses have been published in the
literature, but because of the complexity of the problem, they mostly
take into account one, or at most few, rare decays simultaneously.
Furthermore, most analyses done for general flavor violation in the
MSSM soft terms use the mass insertion approximation (MIA) (see
e.g.~\cite{MIPORO}) which significantly simplifies calculations but
does not produce correct results when flavor violation in the
superpartner sector becomes strong.
 
In a series of papers since 1997~\cite{MIPORO, POROSA, ROS99, BCRS0,
  ROS01, BCRS1, CHRO, BCRS, BEJR, DRT, ROS09}, many supersymmetric
FCNC and CP-violating observables were analyzed with loop-level
accuracy within the setup of the fully general $R$-parity conserving
MSSM without resorting to any MIA-type expansions.  A FORTRAN 77
computer programs based on the common set of Feynman rules of
Ref.~\cite{PRD41} were developed for each process.  Because these
programs use the same conventions, input parameters, and internal data
structures, they can naturally interface with one another.  Combining
these works, we present in this article \code{} - a publicly available
computer code that simultaneously calculates the set of important
$\Delta F=0,1,2$ FCNC and CPV observables in the framework of the
general MSSM.  The current version (1.0) of the program takes a set of
MSSM parameters and calculates the processes\footnote{In the current
  version \code{} calculates also the full one-loop corrections to
  lepton flavor violating $B$-meson decays such as $B\to \mu \tau$.
  However, it is known that contributions to the amplitudes of these
  processes are greatly enhanced at large $\tan\beta$ by formally two
  loop double penguin diagrams~\cite{EDR} which currently are not
  included in the code.  Thus, \code{} can be used to estimate such
  decay rates only at low $\tan\beta \lsim 10$.  } listed in
Table~\ref{tab:proc}.

\begin{table}[t]
\begin{center}
\begin{tabular}{|lcr|}
\hline
Observable & &Experiment \\ \hline
\multicolumn{3}{|c|}{$\Delta F=0$}\\ \hline \hline
$|d_{e}|$(ecm) & &$<1.6 \times 10^{-27}$~\cite{de} \\
$|d_{\mu}|$(ecm) & &$<2.8\times 10^{-19}$~\cite{dmu} \\
$|d_{\tau}|$(ecm) & &$<1.1\times 10^{-17}$~\cite{pdg} \\
$|d_{n}|$(ecm) & &$<2.9 \times 10^{-26}$~\cite{dn} \\ \hline
%
\multicolumn{3}{|c|}{$\Delta F=1$}\\ \hline \hline
$\mathrm{Br}(K_{L }\to \pi^{0} \nu \nu)$ & & $< 6.7\times 10^{-8}$~\cite{Kpinunu} \\
$\mathrm{Br}(K^{+}\to \pi^{+} \nu \nu)$ & & $17.3^{+11.5}_{-10.5}\times
10^{-11}$~\cite{Kpipnunu} \\
$\mathrm{Br}(B_{d}\to e e)$ & & $<1.13\times 10^{-7}$~\cite{bdee}\\
$\mathrm{Br}(B_{d}\to \mu \mu)$ & & $<1.8\times 10^{-8}$~\cite{bmumu}\\
$\mathrm{Br}(B_{d}\to \tau \tau)$ & & $<4.1\times 10^{-3}$~\cite{bdtautau}\\
$\mathrm{Br}(B_{s}\to e e)$ & & $<7.0\times 10^{-5}$~\cite{bsee}\\
$\mathrm{Br}(B_{s}\to \mu \mu)$ & & $<5.8\times 10^{-8}$~\cite{bmumu}\\
$\mathrm{Br}(B_{s}\to \tau \tau)$ & & $--$\\
$\mathrm{Br}(B\to X_{s} \gamma)$ & & $(3.52\pm 0.25)
\times 10^{-4}$~\cite{dmbd}\\ 
\hline
%
\multicolumn{3}{|c|}{$\Delta F=2$}\\ \hline \hline
$|\epsilon_{K}|$ & & $(2.229 \pm 0.010)\times 10^{-3}$~\cite{pdg} \\
$\Delta M_{K}$ & & $(5.292 \pm 0.009)\times 10^{-3}~\mathrm{ps}^{-1}$~\cite{pdg}\\ 
$\Delta M_{D}$ & & $(2.37^{+0.66}_{-0.71})
\times 10^{-2}~\mathrm{ps}^{-1}$~\cite{pdg}\\
$\Delta M_{B_{d}}$ & & $(0.507 \pm 0.005)~\mathrm{ps}^{-1}$~\cite{dmbd}\\  
$\Delta M_{B_{s}}$ & & 
$(17.77 \pm 0.12)~\mathrm{ps}^{-1}$~\cite{dmbs}\\  \hline
\end{tabular}
\end{center}
\label{tab:proc}
\caption{List of observables calculated by \code{} and their currently
  measured values or 95\% C.L bounds (except for $\mathrm{Br}(B_{d}\to
  e e)$ and $\mathrm{Br}(B_{d}\to \tau \tau)$ for which the 90\% C.L
  bounds are given.).}
\end{table}%

%
Several programs allowing to analyze various aspects of the MSSM
flavor phenomenology have been published.  The most relevant to
\code~are: {\tt CPsuperH}\cite{Apostolos}, {\tt SusyBSG}\cite{Pietro}
and {\tt SuperIso}\cite{Mahmoudi}.
{\tt SusyBSG} is dedicated to high-precision predictions for $B\to s
\gamma$ while {\tt CPsuperH} and {\tt SuperIso} calculate processes
similar to the ones computed by \code.  However, these existing codes
are restricted to the Minimal Flavor Violation scenario.  Thus, to the
best of the authors' knowledge, \code{} is the first program which can
simultaneously calculate the set of rare decays listed in
Table~\ref{tab:proc} without any (apart from the $R$-parity
conservation) restrictions on the choice of MSSM parameters.
Other publicly available codes that are relevant to \code{} (which can
e.g.  calculate the MSSM soft parameters used as input to \code, or
for the same set of input parameters calculate non-FCNC related
observables) are {\tt FeynHiggs}\cite{Sven}, {\tt SoftSUSY}\cite{Ben},
{\tt SuSpect}\cite{Suspect}, {\tt SPheno}\cite{Porod}, {\tt
  MicrOMEGAs}\cite{micromega}, {\tt DarkSUSY}\cite{darksusy} and {\tt
  NMHDECAY}\cite{nmhdecay}.

In summary, the basic features of \code{} are:
\begin{itemize}
\item The program utilizes the most general $R$-parity conserving
  Lagrangian for the MSSM.  In addition to standard soft breaking
  terms, it can even accommodate additional non-holomorphic terms,
  such as
\bea
A_d^{'IJ} H_i^{2\star} Q_i^I D^J +
A_u^{'IJ} H_i^{1\star} Q_i^I U^J + \mathrm{H.c.} \;,
\eea
that, for example, do not appear in the minimal supergravity scenario
but are present in the most general softly broken supersymmetric
effective Lagrangian~\cite{Hall}.
\item There is no limit on the size of flavor violating parameters
  because the calculation does not rely on the MIA expansion.  Complex
  ``mass insertions'' of the form
\bea
\delta^{IJ}_{QXY} &=& \frac{(M^2_{Q})^{IJ}_{XY}}{\sqrt{
    (M^2_{Q})^{II}_{XX} (M^2_{Q})^{JJ}_{YY}
}}\;, 
\label{eq:phys:massinsert}
\eea
($I,J$ denote quark flavors, $X,Y$ denote superfield chirality, and
$Q$ indicates either the up or down quark superfield sector, similarly
for slepton superfields) are taken as inputs, but they only serve to
conveniently parametrize the sfermion mass matrices.  \code{}
numerically calculates the exact tree-level spectrum and mixing
matrices, which are later used in loop calculations.

\item As an intermediate step, parton-level form factors for quark and
  lepton 2-, 3- and 4-point Green functions are calculated.  They are
  later dressed in hadronic matrix elements (see Table~\ref{tab:green}
  in Section~\ref{sec:structure}) to obtain predictions for the
  physical quantities listed in Table~\ref{tab:proc}.  The set of
  Green's functions computed by \code{} as intermediate ``building
  blocks'' is quite universal and can be used by other authors to
  calculate other processes.
\item The program runs fairly quickly.  On a Mac PowerBook G4 with GNU
  FORTRAN g77 it returns the output for a single parameter set run
  within a second.
\end{itemize}

We note that the current \code{} version does not resum higher-order
corrections in the limit of large $\tan\beta$.  Such corrections in
the MFV scenario can easily dominate the SM result for $\tan\beta\gsim
30-40$.  However, even in that case new sources of flavor violation
often give comparable or more significant contributions than the
MFV-type $\tan\beta$-enhanced corrections.  Thus, for large
$\tan\beta$ one should perform the resummation of leading higher-order
terms in the presence of the non-vanishing flavor violation in the
sfermion mass matrices.  Unfortunately, such resummation is not yet
fully understood, i.e., although there are theoretical
ideas~\cite{DP,BCRS,CN,Foster} on how to resum these contributions
beyond MFV, it is quite difficult to implement them into \code{}
without losing numerical stability for large mass insertions.  Thus,
even though some parts of \code{} (e.g.  $\bar B B$ mixing and the
$B\ra l^+l^-$ decays) were originally devised to perform a large
$\tan\beta$ resummation for the MSSM parameter choice restricted to
the MFV case, in the present version this option has been deactivated
for consistency.  We hope to improve this in future versions of the
code.  The current version of \code{} should be used for low to
moderate values of $\tan\beta$ ($\tan\beta\lsim 30$), however even in
the case of significant supersymmetric flavor violation it should
still produce reasonably accurate results even for $\tan\beta$ beyond
this range.

The rest of the paper is organized as follows.  In
Section~\ref{sec:mssmlag} we define the general structure of the MSSM
Lagrangian following Ref.~\cite{PRD41} to facilitate comparison of the
conventions used in \code{} with others used in the literature and to
connect the variables used in the code with physical quantities.
Section~\ref{sec:structure} describes the internal structure of
\code{}, the most important steps of calculations, and the file
structure of the library.  In Section~\ref{sec:init} we carefully
present the initialization sequence for \code{}, defining input
parameters and how they are used.  Routines for calculating the FCNC
and CPV observables collected in Table~\ref{tab:proc} are described in
Section~\ref{sec:proc}.  We conclude with a summary of the
presentation.  Appendix~\ref{app:inst} contains brief instructions on
how to install and run the \code{} package.
In Appendices~\ref{app:code} and~\ref{app:infile} we provide templates
for initializing \code{} from within the program and using an external
file in the SLHA2 format~\cite{SLHA2}, respectively. Both of these
templates produce the set of test results listed in
Appendix~\ref{app:output}.

\code{} can be downloaded from the following address:

\vspace*{0.3cm} \centerline{\webpage}

\section{Lagrangian, conventions and the tree level masses}
\label{sec:mssmlag}

\subsection{Lagrangian parameters}
\label{sec:parlist}

\code{} follows the conventions for the MSSM Lagrangian and Feynman
rules for the most general $R$-parity conserving version of the MSSM
Lagrangian given in~\cite{PRD41}.  Over 100 Lagrangian parameters are
taken as input to \code{} and can be initialized independently.

For completeness and for easier comparison with conventions used in
other sources, we present here the full list of the general MSSM
couplings.  They can be classified by sectors of the theory:

\noindent {\bf 1.  Gauge sector.} $g_1$, $g_2$, $g_3$ denote the
coupling constants of gauge groups $U(1)_Y$, $SU(2)_L$, $SU(3)_c$,
respectively.

\noindent {\bf 2.  Superpotential and Yukawa couplings.} The
superpotential and the soft breaking sfermion couplings are written
after the rotations of superfields to the super-KM basis in which the
Yukawa couplings are diagonal and the soft parameters are redefined
accordingly to account for accommodate these flavor rotations (see
e.g.~\cite{MIPORO}).

We do not assume the existence of the heavy right-handed
neutrino/sneutrino supermultiplet and neglect related terms in the
Lagrangian.\footnote{The modifications to the phenomenology of the
  MSSM from the presence of a heavy right neutrino supermultiplet are
  discussed in~\cite{DEHARO}.  Some numerical codes concerning the
  problem can be obtained from its authors.}  Then the most general
form of the $R$-parity conserving MSSM superpotential takes the form:
\bea
W = \mu\epsilon_{ij} H^1_i H^2_j + \epsilon_{ij} Y_l^{I} H^1_i L_j^I
R^I + \epsilon_{ij} Y_d^{I} H^1_i Q_j^I D^I + \epsilon_{ij} Y_u^{I}
H^2_i Q_j^I U^I.
\eea
Capital indices ($I, J, K \ldots$) label matter field generations and
run from $1$ to $3$.  Lower-case indices ($i,j,\ldots$) are $SU(2)_L$
indices (we use $\epsilon_{12}=-1$).  $SU(3)_c$ indices are not
written explicitly; we assume that the $Q$ supermultiplets are QCD
triplets and the $D$ and $U$ supermultiplets are anti-triplets.
At tree level, quark and lepton masses are related to Yukawa couplings
by (note that $Y_l^I, Y_d^I$ are negative in our convention):
\bea
m_e^I = - {v_1 Y_l^I \over \sqrt{2}} \;, \hskip 2cm m_d^I = - {v_1
  Y_d^I\over \sqrt{2}} \;, \hskip 2cm m_u^I = {v_2 Y_u^I\over \sqrt{2}}\;.
\eea
It follows that in \code{} the fermion masses and the elements of the
Cabibbo-Kobayashi-Maskawa (CKM) matrix, rather than the Yukawa
couplings, are used as input parameters.

\noindent {\bf 3.  Soft gaugino mass terms} for the $U(1)_Y$,
$SU(2)_L$ and $SU(3)_c$ gauge groups
\bea
\frac{1}{2} M_1 \lambda_B \lambda_B + \frac{1}{2} M_2 \lambda_A^i
\lambda_A^i + \frac{1}{2} M_3 \lambda_G^a \lambda_G^a + \mathrm{h.c.}
\eea

\noindent {\bf 4.  Soft-breaking mass terms for the scalar fields.}
\bea
& - m_{H_1}^2 H_i^{1\star} H_i^1 - m_{H_2}^2 H_i^{2\star} H_i^2 -
(m_L^2)^{IJ} L_i^{I\star} L_i^J - (m_E^2)^{IJ} E^{I\star} E^J&
\nonumber\\
& - (K m_Q^2 K^\dagger)^{IJ} Q_1^{I\star} Q_1^J - (m_Q^2)^{IJ}
Q_2^{I\star} Q_2^J - (m_D^2)^{IJ} D^{I\star} D^J - (m_U^2)^{IJ}
U^{I\star} U^J & \;.
\eea

\noindent {\bf 5.  Trilinear scalar couplings corresponding to
  superpotential Yukawa terms.}
\bea
m_{12}^2 \epsilon_{ij} H_i^1 H_j^2 + \epsilon_{ij} A_l^{IJ} H_i^1
L_j^I E^J + \epsilon_{ij} A_d^{IJ} H_i^1 Q_j^I D^J + \epsilon_{ij}
A_u^{IJ} H_i^2 Q_j^I U^J + \mathrm{h.c.}
\eea

\noindent {\bf 6.  Non-standard trilinear scalar couplings} involving
complex conjugated Higgs fields (sometimes called ``non-analytic
terms'').
\bea
A_l^{'IJ} H_i^{2\star} L_i^I E^J + A_d^{'IJ} H_i^{2\star} Q_i^I D^J +
A_u^{'IJ} H_i^{1\star} Q_i^I U^J + \mathrm{h.c.}
\eea
Usually these couplings are not considered as they are not generated
in standard SUSY breaking models.  However, for completeness they are
included in~\code{} and by default initialized to zero.  Users of
\code{} may decide to set them to some non-vanishing values in order
to check their impact on rare decay phenomenology.

In general, the mass parameters $\mu$, $m_{12}^2$, $m_Q^2$,
$M_{1,2,3}$, and the trilinear soft couplings may be complex.  Global
rephasing of all fermion fields of the theory and of one of the Higgs
multiplets can render two of these parameters real~\cite{POROSA}.  We
choose them to be the gluino mass $M_3$ and the soft Higgs mixing term
$m_{12}^2$.  The latter choice keeps the Higgs vacuum expectation
values (VEV) and, therefore, the parameter $\tan\beta$ real at tree
level.

\subsection{Physical tree level masses and mixing angles}
\label{sec:physmass}

Mass matrices of the MSSM particles can be written in terms of the
parameters of Section~\ref{sec:parlist}.  In \code,
following~\cite{PRD41}, we consistently use matrix notation for all
fields, including neutral and charged Higgs bosons.  Such a notation
simplifies the expressions for loop calculations.  In this Section we
explicitly write down all mass matrices to fix our sign conventions
relative to other choices in the literature.

\noindent {\bf 1. Higgs sector.}  We denote the CP-even and CP-odd
neutral scalars as $H^0_i$ and $A^0_i$, respectively, with $i=1,2$.
In terms of more common notation, $(H^0_1,H^0_2)\equiv (H^0,h^0)$ and
$(A^0_1,A^0_2)\equiv (A^0,G^0)$.  These are related to the initial
Higgs doublets by (no sum over~$i$):
\bea
\real H_i^i &=&  \frac{1}{\sqrt{2}} (Z_R^{ij} H_j^0 + v_i)\;,\nonumber\\
\imag H_i^i &=& \frac{1}{\sqrt{2}} Z_H^{ij} A_j^0\;.
\eea
In these formulae $v_1,v_2$ are the VEVs of the two neutral components
of the Higgs doublets and $Z_R, Z_H$ are the mixing matrices in the
CP-even and CP-odd Higgs sectors, respectively.

The mixing matrix $Z_R$ and the masses of $H_i^0$ can be obtained by
diagonalizing the CP-even Higgs mass$^2$ matrix:
\bea
Z_R^T 
\left(
\begin{array}{cc}
- m_{12}^2 {v_2\over v_1} + {e^2 v_1^2 \over 4 s_W^2 c_W^2} 
& 
m_{12}^2 - {e^2 v_1 v_2 \over 4 s_W^2 c_W^2} 
\\ 
m_{12}^2 - {e^2 v_1 v_2 \over 4 s_W^2 c_W^2} 
& 
- m_{12}^2 {v_1\over v_2} + {e^2 v_2^2 \over 4 s_W^2 c_W^2}
\end{array}
\right)
Z_R =  \left(
\begin{array}{cc}
  M^2_{H^0_1}  &0\\
  0 & M^2_{H^0_2}    
\end{array}
\right)
\equiv  \left(
\begin{array}{cc}
  M^2_H  &0\\
  0 & M^2_h    
\end{array}
\right).  
\eea
$A_1^0 (\equiv A^0)$ has mass $M_A^2 = m_{H_1}^2 + m_{H_2}^2 +
2|\mu|^2$.  \code{} assumes $R_\xi$ gauge with $\xi=1$, so the neutral
Goldstone boson $A_2^0 (\equiv G^0)$ has the mass $M_{G^0} = M_Z$.

The mixing matrices $Z_H, Z_R$ are parametrized as follows:
\bea
Z_H = \left(
  \begin{array}{cc}
    {\sin}{\beta}&-  {\cos}{\beta}\\
    {\cos}{\beta}&  {\sin}{\beta}
  \end{array}
\right)\;,
\hskip 1cm
Z_R = \left(
\begin{array}{cc}
  {\cos}{\alpha}&-  {\sin}{\alpha}\\
  {\sin}{\alpha}&  {\cos}{\alpha}
\end{array}
\right)\;,
\eea
with the angles $\alpha$ and $\beta$ determined by
\bea
\tan \beta = {v_2 \over v_1} \;, &\hskip
2cm& 0\leq \beta \leq  \frac{\pi}{2}\;, \nonumber\\[3mm]
\tan 2\alpha = \tan 2\beta\, {M^2_A + M_Z^2 \over M_A^2 - M_Z^2}\;, &\hskip
2cm& -\frac{\pi}{2}\leq \alpha \leq 0.
\eea
Charged Higgs scalars are denoted by $H^\pm_i\equiv H^\pm,G^\pm$ and
are related to the initial Higgs doublet again by the matrix $Z_H$:
\bea
\left(
  \begin{array}{c}
  H_2^{1\star}  \\
  H_1^2
\end{array}
\right) = Z_H  \left(
\begin{array}{c}
  H_1^+  \\
  H_2^+  
\end{array}
\right).
\eea
The physical charged Higgs boson has mass
\bea
M_{H_1^{\pm}}^2 = M_W^{2} + m_{H_1}^2 + m_{H_2}^2 + 2|\mu|^2,
\eea
while the charged Goldstone bosons $G^\pm$ have masses
$M_{G^\pm}=M_W$.

\noindent {\bf 2.  Gaugino sector.}  The chargino masses and mixing
matrices $Z_+$ and $Z_-$ are defined by the relation
\bea
(Z_-)^T 
\left(
\begin{array}{cc}
 M_2 & {ev_2 \over \sqrt{2}s_W}
\\  {ev_1 \over \sqrt{2}s_W} & \mu
\end{array}
\right)
Z_+ =  \left(
\begin{array}{cc}
  M_{\chi_1} & 0
\\
  0 & M_{\chi_2}  
\end{array}
\right).
\eea
In \code{} we choose $Z_-,Z_+$ such that both masses $M_{\chi_i}$ are
real positive and $M_{\chi_2} > M_{\chi_1}$.

\noindent The neutralino tree level masses are given by
\bea
Z_N^T
\left(
\begin{array}{cccc}
  M_1 & 0 & {-ev_{1} \over 2c_W} & {ev_{2} \over
    2c_W} \\ 
  0 & M_2 & {ev_{1} \over 2s_W} & {-ev_{2} \over 
    2s_W} \\ 
  {-ev_{1} \over 2c_W} & {ev_{1} \over 2s_W} & 0
  & -\mu \\ 
  {ev_{2} \over 2c_W} & {-ev_{2} \over 2s_W} &-\mu
  & 0
\end{array}
\right)
Z_N &=& \left(
\begin{array}{ccc}
  M_{\chi^0_1}  &&0\\
  &{\ddots}&\\
  0&& M_{\chi^0_4} 
\end{array}
\right),
\eea
where again we use the ambiguity in the definition of the $Z_N$ matrix
to choose to make all $M_{\chi^0_i}$ real positive and increasingly
ordered.

\noindent {\bf 3.  Slepton sector.}  The three complex sneutrino
fields have tree level masses and the mixing matrix $Z_{\tilde\nu}$
defined by:
\bea
Z_{\tilde\nu}^{\dagger} \left({e^2(v^2_1-v^2_2) \over 8 s_W^2 c_W^2}
  \hat 1 + m_L^2\right) Z_{\tilde\nu} &=& \left(
\begin{array}{ccc}
  M_{\tilde\nu_1}^2 &&0\\
  &{\ddots}&\\
  0&&M_{\tilde\nu_3}^2 
\end{array}
\right).
\eea
The mass matrix for the six charged sleptons
\bea
{\cal M}^2_L = \left(
\begin{array}{cc}
  {e^2(v_1^2-v_2^2)(1-2c_W^2) \over 8 s_W^2 c_W^2} 
  + {v_1^2 Y_l^2\over 2} + (m_L^2)^T & 
  {v_2\over \sqrt{2}} (Y_l \mu^{\star} - A_l^{'}) + {v_1\over \sqrt{2}} A_l \\
  {v_2\over \sqrt{2}}  (Y_l \mu - A_l^{'\dagger}) 
  + {v_1\over \sqrt{2}} A_l^\dagger & 
  - {e^2(v_1^2-v_2^2) \over 4c_W^2} + {v_1^2 Y_l^2\over 2} + m_E^2
\end{array}
\right)
\label{eq:slmass}
\eea
is diagonalized by the unitary matrix $Z_L$,
\bea
Z_L^{\dagger} {\cal M}^2_L Z_L = \left(
\begin{array}{ccc}
  M_{L_1}^2 &&0\\
  &{\ddots}&\\
  0&&M_{L_6}^2 
\end{array}
\right).
\label{eq:sldiag}
\eea

\noindent {\bf 4.  Squark sector.}  Analogously, for the up and down
squarks one has:
\bea
{\cal M}_U^2 = \left(
\begin{array}{cc}
  {e^2(v_1^2-v_2^2)(4c_W^2-1) \over 24 s_W^2 c_W^2} 
  + {v_2^2 Y_u^2 \over 2} + (Km_Q^2K^\dagger)^T & 
  -{v_1\over \sqrt{2}} (Y_u \mu^{\star} + A_u^{'}) - {v_2\over \sqrt{2}} A_u \\
  -{v_1\over \sqrt{2}} (Y_u \mu + A_u^{'\dagger}) 
  - {v_2\over \sqrt{2}} A_u^\dagger &
  {e^2(v_1^2-v_2^2) \over 6c_W^2} + {v_2^2 Y_u^2 \over 2} + m_U^2
\end{array}
\right) \;,
\label{eq:sumass}
\eea
\bea
Z_U^T {\cal M}_U^2 Z_U^{\star} = \left(
\begin{array}{ccc}
  M_{U_1}^2 &&0\\
  &{\ddots}&\\
  0&&M_{U_6}^2 
\end{array}
\right)\;.
\label{eq:sudiag}
\eea
\bea
{\cal M}_D^2 = \left(
\begin{array}{cc}
  - {e^2(v_1^2-v_2^2)(1+2c_W^2) \over 24 s_W^2 c_W^2} 
  + {v_1^2 Y_d^2 \over 2} + (m_Q^2)^T & 
  {v_2\over \sqrt{2}} (Y_d \mu^{\star} - A_d^{'}) + {v_1\over \sqrt{2}} A_d \\
  {v_2\over \sqrt{2}} (Y_d \mu - A_d^{'\dagger}) 
  + {v_1\over \sqrt{2}} A_d^\dagger &
  - {e^2(v_1^2-v_2^2) \over 12c_W^2} + {v_1^2 Y_d^2 \over 2} + m_D^2
\end{array}
\right)\;,
\label{eq:sdmass}
\eea
\bea
Z_D^{\dagger} {\cal M}_D^2 Z_D = \left(
\begin{array}{ccc}
  M_{D_1}^2 &&0\\
  &{\ddots}&\\
  0&&M_{D_6}^2 
\end{array}
\right).
\label{eq:sddiag}
\eea
Note that $Z_U$ is defined with a complex conjugate compared to the
definitions of $Z_L$ and $Z_D$.  Thus all positively charged sfermion
mass eigenstates are multiplied by $Z_X^{ij}$, while negatively
charged eigenstates are multiplied by $Z_X^{ij\star}$.

\subsection{Interfacing with the Les Houches Accord}
\label{sec:slha}

\code{} has been in development since 1996, long before the Les
Houches Accord~\cite{SLHA} (SLHA) for common MSSM Lagrangian
conventions was agreed.  Because of that, it was not feasible to
change the internal \code{} structure as it would require careful
checking and rewriting of thousands of lines of a complicated code.
Therefore we have decided to keep the conventions of~\cite{PRD41} for
the internal calculations in \code{}.  In Table~\ref{tab:slha} we
summarize the differences of our conventions and those of the latest
extended SLHA 2~\cite{SLHA2}.  These differences are quite minor and
translation can be done by changing few signs and/or transposing
matrices in the soft SUSY breaking sector.  Thus, for the input
parameters of \code{} we leave the choice of convention as a
user-defined option.

\begin{table}[htbp]
\begin{center}
\begin{tabular}{|c|c|}
\hline
SLHA 2~\cite{SLHA2} & Ref.~\cite{PRD41}\\
\hline 
& \\[-4mm]
$\hat T_U$, $\hat T_D$, $\hat T_E$ & $-A_u^T$, $+A_d^T$, $+A_l^T$\\
$\hat m_{\tilde Q}^2$, $\hat m_{\tilde L}^2$ & $m_Q^2$, $m_L^2$ \\
$\hat m_{\tilde u}^2$, $\hat m_{\tilde d}^2$, $\hat m_{\tilde l}^2$ &
$(m_U^2)^T$, $(m_D^2)^T$, $(m_E^2)^T$ \\
${\cal M}_{\tilde u}^2$, ${\cal M}_{\tilde d}^2$ & $({\cal M}_U^2)^T$,
$({\cal M}_D^2)^T$ \\[1mm]
\hline
\end{tabular}
\end{center}
\caption{Comparison of~SLHA~\cite{SLHA2} and Ref.~\cite{PRD41}
  conventions.}
\label{tab:slha}
\end{table}

Currently \code{} does not use the super-PMNS basis for the lepton and
slepton sector; only the charged lepton Yukawa matrix (and not the
neutrino mass matrix) is diagonalized.  The super-PMNS basis can
become helpful once new experiments are able to identify the flavor of
the neutrinos produced in rare decays, but at present this is not
experimentally feasible.

\section{Structure of the code}
\label{sec:structure}

Calculations in \code{} take the following steps:

\noindent {\bf 1.  Parameter initialization.}  This is the most
important step for \code{} users and is described in detail
Section~\ref{sec:init}.  Users can adjust the basic Standard Model
parameters according to latest experimental data and initialize all
(or the chosen subset of) supersymmetric soft masses and couplings and
Higgs sector parameters listed in Section~\ref{sec:parlist}.

\noindent {\bf 2.  Calculation of the physical masses and the mixing
  angles.}  After setting the input parameters, \code{} calculates the
eigenvalues of the mass matrices of all MSSM particles and their
mixing matrices at the tree level.  Diagonalization is done
numerically without any approximations.

\noindent {\bf 3.  Calculation of Wilson coefficients at the SUSY
  scale}.  Physical tree-level masses and mixing matrices are used to
evaluate exact one-loop Wilson coefficients of the effective operators
required for a given process.  Again, the formulae used in the code
are exact, i.e.  do not rely on any approximations, such as the MIA
expansion.  In the current version, \code{} calculates Wilson
coefficients generated by the diagrams listed in
Table~\ref{tab:green}.  All Wilson coefficients are calculated at the
high energy scale, assumed to be the average mass of SUSY particles
contributing to a given process or the top quark scale.

\begin{table}[htbp]
\label{tab:green}
\begin{center}
\begin{tabular}{|cp{1mm}|p{1mm}cp{1mm}|p{1mm}c|}
\hline
Box &&& Penguin &&& Self energy \\[2mm] \hline\hline
$dddd$ &&& $Z\bar d d$, $\gamma \bar d d$, $g \bar d d$ &&& $d$-quark \\
$uuuu$ &&& $H_i^0 \bar d d$, $A_i^0 \bar d d$ &&& $u$-quark \\
$ddll$ &&& $H_i^0 \bar u u$, $A_i^0 \bar u u$ &&& \\
$dd\nu\nu$ &&& &&& \\
\hline
\end{tabular}\\[2mm]
\caption{One loop parton level diagrams implemented in \code{}.}
\end{center}
\end{table}

It is important to stress that \code{} accepts fermion generation
indices and Higgs boson indices as input parameters.  Thus in
Table~\ref{tab:green} $d$ and $u$, $l$ and $\nu$ denote quarks or
leptons of {\em any} generation and, similarly, $H_i^0$ and $A_i^0$
denote {\em any} type of the neutral Higgs bosons.  Hence, the actual
number of amplitudes which can be calculated using combinations of
these form factors is much larger than needed for the rare decay rates
currently implemented fully in \code{}. We plan to add new processes
in future releases of our library.

\noindent {\bf 4.  Strong corrections.}  In its final step \code{}
performs (when necessary) the QCD evolution of Wilson coefficients
from the high energy (SUSY or top quark mass) scale to the low energy
scale appropriate for a given rare decay, calculates the relevant
hadronic matrix elements, and outputs predictions for physical
quantities.  The formulae for QCD and hadronic corrections are
primarily based on calculations performed in the SM and supplemented,
when necessary, with contributions from non-standard operators which
usually are neglected in the SM, because they are suppressed by powers
of the light quark Yukawa couplings.  This part of \code{} is based on
analyses published by other authors, whereas points 1-3 are
implemented using our own calculations.  The accuracy of strong
corrections differ from process to process, from negligible or small
(leptonic EDM, ``gold-plated'' decay modes $K\ra \pi\bar \nu
\nu$~\cite{BB98}) to order of magnitude uncertainties (unknown long
distance contributions to $\Delta m_K$ or $\Delta m_D$).  Even in the
case of large QCD uncertainties, the result of the calculation
performed by \code{} can be of some use.  Flavor violation in the
sfermion sector can lead to huge modifications of many observables,
sometimes by several orders of magnitude, so that comparison with
experimental data can help to constrain the soft flavor-violating
terms even if strong corrections are not very well known.

Below we list the files included in the \code{} library with a brief
description of their content and purpose.
\begin{itemize}
\item[{\tt eisch1.f:}] auxiliary numerical routine - hermitian matrix
  diagonalization
\item[{\tt vegas.f:}] auxiliary numerical routine - Vegas Monte Carlo
  integration
\item[{\tt rombint.f:}] auxiliary numerical routine - Romberg
  numerical integration
\item[{\tt sflav\_io.f:}] input routine for reading of the SLHA2
  format; test output routines
\item[{\tt b\_fun.f:}] general 2-point loop functions
\item[{\tt db\_fun.f:}] derivatives of general 2-point loop functions
\item[{\tt c\_fun.f:}] general 3-point loop functions 
\item[{\tt cd\_fun.f:}] 3-, 4- and some 5-point loop functions at
  vanishing external momenta
\item[{\tt vh\_def.f:}] definitions of Higgs boson tree-level vertices
\item[{\tt vg\_def.f:}] definitions of gauge boson tree-level vertices
\item[{\tt vf\_def.f}] definitions of fermion tree-level vertices
\item[{\tt mh\_init.f:}] initialization of MSSM parameters 
\item[{\tt mh\_diag.f:}] diagonalization of tree level mass matrices;
  outputs physical masses and mixing angles
\item[{\tt qcd\_fun.f:}] auxiliary QCD calculations - running
  $\alpha_s$, running quark masses etc.
\item[{\tt d\_self0.f:}] $d$-quark self-energy
\item[{\tt u\_self0.f:}] $u$-quark self-energy
\item[{\tt sff\_fun0.f:}] form factors of the general
  scalar-fermion-fermion 1-loop triangle diagram
\item[{\tt sdd\_vert0.f:}] CP-even Higgs-$d$ quark-$d$ quark 1-loop
  triangle diagram
\item[{\tt pdd\_vert0.f:}] CP-odd Higgs-$d$ quark-$d$ quark 1-loop
  triangle diagram
\item[{\tt suu\_vert0.f:}] CP-even Higgs-$u$ quark-$u$ quark 1-loop
  triangle diagram
\item[{\tt puu\_vert0.f:}] CP-odd Higgs-$u$ quark-$u$ quark 1-loop
  triangle diagram
\item[{\tt zdd\_vert0.f:}] $Z$ boson-$d$ quark-$d$ quark 1-loop
  triangle diagram
\item[{\tt ddg\_fun.f:}] form factors for the general gauge
  boson-fermion-fermion 1-loop triangle diagram
\item[{\tt dd\_gluon.f:}] $d$ quark-$d$ quark-gluon 1-loop triangle
  diagram
\item[{\tt dd\_gamma.f:}] $d$ quark-$d$ quark-photon 1-loop triangle
  diagram
\item[{\tt bsg\_nl.f:}] formulae for $\mathrm{Br}(B\ra X_s \gamma)$, including
  QCD corrections
\item[{\tt dd\_ll.f:}] $d$ quark-$d$ quark-lepton-lepton 1-loop
  box diagram
\item[{\tt dd\_vv.f:}] $d$ quark-$d$ quark-neutrino-neutrino 1-loop
  box diagram
\item[{\tt phen\_2q.f:}] formulae for $Br(K_L^0\ra \pi^0 \bar\nu
  \nu)$, $\mathrm{Br}(K^+\ra \pi^+ \bar\nu \nu)$ and $\mathrm{Br}(B_{s(d)}\ra l^+l^-)$
  including QCD corrections and hadronic matrix elements
\item[{\tt dd\_mix.f:}] 4-$d$ quark 1-loop box diagram
\item[{\tt uu\_mix.f:}] 4-$u$ quark 1-loop box diagram
\item[{\tt phen\_4q.f:}] formulae for the meson mixing observables:
  $\Delta m_K$, $\epsilon_K$, $\Delta m_D$, $\Delta m_{B_{d(s)}}$
  including QCD corrections and hadronic matrix elements
\item[{\tt edm\_l.f:}] lepton electric dipole moment 
\item[{\tt cdm\_d.f:}] $d$-quark chromoelectric dipole moment 
\item[{\tt cdm\_u.f:}] $u$-quark chromoelectric dipole moment 
\item[{\tt cdm\_g.f:}] gluon chromoelectric dipole moment 
\item[{\tt edm\_d.f:}] $d$-quark electric dipole moment 
\item[{\tt edm\_u.f:}] $u$-quark electric dipole moment 
\item[{\tt edm\_n.f:}] $u$-quark electric dipole moment 
\end{itemize}
All the 2-, 3- and 4-point Green functions are calculated for
vanishing external momenta.  As mentioned before, by ``$u$ quark'' and
``$d$ quark'' we mean all generations of quarks.

In addition to the files listed above, the library contains the master
driver file {\tt susy\_flavor.f} which illustrates the proper
initialization sequence for \code{} parameters and produces a set of
test results for the implemented observables.

\section{Parameter initialization in \code}
\label{sec:init}

We now list the input parameters used by \code{}.  These are not
always directly the MSSM Lagrangian parameters given in
Section~\ref{sec:parlist} -- for example, instead of using the $\mu$
parameter and the soft Higgs masses $m_{H_i}^2, m_{12}^2$, it is
customary to use $\tan\beta$ and the CP-odd Higgs mass $M_A$ to
parametrize MSSM Higgs sector.  In its first step, \code{} restores
the Lagrangian parameters of Section ~\ref{sec:parlist} for the given
set of more human-friendly input parameters.  Then, the remaining
routines use the ``raw'' Lagrangian parameters---if necessary they can
also be directly modified by (experienced!) users.

In the rest of this section we describe step-by-step the basic
initialization routines used by \code{}, their arguments and, when
necessary, the FORTRAN common blocks storing the most important data
(other common blocks serve for the internal purposes and usually do
not need to be accessed by users).

By default, \code{} uses the following implicit type declaration in
all routines:\\[2mm]
{\tt implicit double precision (a-h,o-z)}\\[2mm]
so that all variables in \code{} with the names starting from {\tt a}
to {\tt h} and from {\tt o} to {\tt z} are automatically defined as
{\tt double precision} and those with names starting from {\tt i} to
{\tt n} are of {\tt integer} type.  In what follows we explicitly
indicate variables that do not obey this rule.  Such variables are
always listed in explicit type statements inside the procedures.
Complex parameters mentioned in this article are declared in \code{}
as {\tt double complex} type.  Mass parameters are always given in
GeV.

\code{} provides two ways of initializing input parameters.  As the
first option, they can be read from the file {\tt susy\_flavor.in}.
The structure of this file follows the SLHA2 convention~\cite{SLHA2},
with some extensions which we describe in Section~\ref{sec:slhainp}.
Initializing parameters in the input file is simple, it is done by a
call to single subroutine {\tt sflav\_input} and does not require
detailed knowledge of the program internal structure. This option is
particularly convenient for testing a single parameter set but can be
cumbersome for scans over the MSSM parameter space.  Therefore, as a
second option, \code{} also provides a set of routines designed to
initialize parameters defined in the program, which can easily be used
to prepare programs that scan over large parameter sets.  As described
in Section~\ref{sec:proginp}, these routines require more care in use,
as they should be initialized in the proper order, i.e.  the gauge
sector first, then the fermion sector, Higgs sector, and SUSY sectors
at the end (the initialization sequences for the gaugino, slepton and
squark sectors are independent).

An example of a full initialization sequence for \code, illustrating
both options mentioned above, is presented Appendix~\ref{app:code}.
The sample input file {\tt susy\_flavor.in} is given in
Appendix~\ref{app:infile}.  Test output generated for parameters used
in Appendices~\ref{app:code} and~\ref{app:infile} is enclosed in
Appendix~\ref{app:output}.

\subsection{Parameter initialization from the input file}
\label{sec:slhainp}

Input parameters for \code{} can be set by editing appropriate entries
of the file {\tt susy\_flavor.in} and subsequently calling the
subroutine {\tt sflav\_input}, which reads the input file, stores the
the MSSM Lagrangian parameters in FORTRAN common blocks and calculates
tree-level physical masses and mixing matrices.  After calling {\tt
  sflav\_input}, all physical observable described in
Section~\ref{sec:proc} can be calculated.

The input file {\tt susy\_flavor.in} is written in the SLHA2 format,
with some extensions which we list below (for an example of a complete
input file see Appendix~\ref{app:infile}).

1.  We define a non-standard {\tt Block SOFTINP}.  Currently it
contains two control variables, {\tt iconv} and {\tt
  input\_type}. These serve to choose input conventions in the
sfermion sector (in other sectors SLHA2 and Ref.~\cite{PRD41}
agree).\\[2mm]
\begin{tabular}{lcp{125mm}}
Variable value && Sfermion sector parametrization \\[2mm]
${\tt iconv}=1$ && MSSM parameters defined in SLHA 2 conventions.\\
${\tt iconv}=2$ && MSSM parameters defined in conventions of
Ref.~\cite{PRD41}.\\
${\tt input\_type}=1$ && sfermion diagonal trilinear mixing terms
given as dimensionless parameters; all off-diagonal soft terms are
given as dimensionless mass insertions---see comments below on the
data blocks defining the sfermion soft terms.\\
${\tt input\_type}=2$ && sfermion soft terms given as absolute values
of dimension mass$^2$.\\
\end{tabular}

2. \code{} uses the $W$ boson mass as a basic parameter rather than
Fermi constant $G_F$. Therefore, in {\tt susy\_flavor.in} we use 
entry 30 of {\tt Block SMINPUTS} (not used in the standard SLHA2) to
define $M_W$.

3. We allow complex values for $\mu$ and two of the gaugino
masses---chosen to be the $U(1)$ and $SU(2)$ mass terms $M_1$ and
$M_2$.  Their real and imaginary parts are defined in blocks {\tt
  EXTPAR} and {\tt IMEXTPAR}.  We use $\tan\beta$ and the CP-odd Higgs
mass $M_A$ as the input parameters for the Higgs sector.

4. Following the SLHA2 convention, we only define the upper triangle
of each of the hermitian sfermion soft mass matrices in the {\tt
  MSL2IN, MSE2IN, MSQ2IN, MSD2IN, MSU2IN} and {\tt IMMSL2IN, IMMSE2IN,
  IMMSQ2IN, IMMSD2IN, IMMSU2IN} blocks. It is obligatory to define all
entries, both diagonal and upper off-diagonal, since \code{} does not
read diagonal sfermion masses from the {\tt EXTPAR} block. The {\tt
  iconv} parameter defined in the {\tt SOFTINP} block determines if
sfermion parameters are given in SLHA2 or Ref.~\cite{PRD41}
conventions (see Table~\ref{tab:slha}).  Finally, the {\tt
  input\_type} parameter in the {\tt SOFTINP} block defines the format
of the off-diagonal mass terms. If ${\tt input\_type}=1$, the
off-diagonal entries given in {\tt susy\_flavor.in} are assumed to be
dimensionless mass insertions and the flavor violating sfermion mass
terms are calculated as
\bea
(m_X^2)_{IJ} &=& (m_X^2)_{JI}^\star = \delta^{IJ}_{X}
\sqrt{(m_X^2)_{II} (m_X^2)_{JJ}}\; , 
\label{eq:midef}
\eea
where $X=L,E,Q,U,D$ and $I,J$ enumerate superpartners of the
mass-eigenstate quarks.

5. The blocks {\tt TEIN, TDIN, TUIN} and {\tt IMTEIN, IMTDIN, IMTUIN}
define the trilinear sfermion mixing matrices which are generally
non-hermitian. One is required to define all entries. As for the soft
mass terms, the {\tt iconv} parameter chooses the input convention,
SLHA2 or Ref.~\cite{PRD41}.  For the trilinear mixing, the parameter
{\tt input\_type} defines the format and dimension of both the
diagonal and off-diagonal terms.  If ${\tt input\_type}=1$, then all
relevant {\tt susy\_flavor.in} entries are treated as dimensionless
numbers and expanded to full trilinear mixing matrices using
eqs.~(\ref{eq:lrdef},\ref{eq:lrmidef}).  For the diagonal LR terms,
\code{} uses the formulae
\bea
A_l^{II} &=& Y_l^I \left((m_L^2)_{II} (m_E^2)_{II}\right)^{1/4}
a_l^I\;,\nonumber\\
A_d^{II} &=& Y_d^I \left((m_Q^2)_{II} (m_D^2)_{II}\right)^{1/4}
a_d^I\;,\nonumber\\
A_u^{II} &=& Y_u^I \left((m_Q^2)_{II} (m_U^2)_{II}\right)^{1/4}
a_u^I\;,
\label{eq:lrdef}
\eea
where $a_l^I, a_d^I, a_u^I$ are the diagonal trilinear mixing terms
read from the input file.

For the off-diagonal LR terms, \code{} uses
\bea
A_l^{IJ} &=& \delta^{IJ}_{LLR} \sqrt{2}/v_1 \sqrt{(m_L^2)_{II}
  (m_E^2)_{JJ}}\; ,\nonumber\\
A_d^{IJ} &=& \delta^{IJ}_{DLR} \sqrt{2}/v_1 \sqrt{(m_Q^2)_{II}
  (m_D^2)_{JJ}}\; ,\nonumber\\
A_u^{IJ} &=& \delta^{IJ}_{ULR} \sqrt{2}/v_2 \sqrt{(m_Q^2)_{II}
  (m_U^2)_{JJ}}\; .
\label{eq:lrmidef}
\eea
Note that in eqs.~(\ref{eq:lrdef},\ref{eq:lrmidef}) for simplicity we
use $(m_Q^2)_{II}$ as the diagonal mass scale for both up and down
left squark fields (in general related by the CKM rotation, see
eqs.~(\ref{eq:sumass},\ref{eq:sdmass})).

\subsection{Parameter initialization inside the program}
\label{sec:proginp}

\code{} input parameters can also be initialized directly inside the
driver program using the set of routines described below.  Before the
proper initialization sequence, the user can set the {\tt iconv}
variable value to choose the input convention:\\[2mm]
\begin{tabular}{lp{1mm}l}
{\tt common/sf\_cont/eps,indx(3,3),iconv} \\
\hskip 11mm {\tt iconv=1} && SLHA2~\cite{SLHA2} input conventions \\
\hskip 11mm {\tt iconv=2} && \cite{PRD41} input conventions\\
\end{tabular}\\[2mm]
After choosing the input conventions, one should subsequently
initialize the gauge, matter fermion, Higgs, SUSY fermion and sfermion
sectors, using the procedures described in detail in the following
sections.

\subsubsection{Gauge sector}

As input, \code{} takes the gauge boson masses ($M_W, M_Z$) and the
gauge coupling constants (electromagnetic and strong) at the $M_Z$
scale.  They are initialized by:\\[2mm]
\begin{tabular}{lp{1mm}p{95mm}}
 Routine and arguments && Purpose and MSSM parameters \\[2mm]
{\tt vpar\_update(zm,wm,alpha\_{em})} && Sets electromagnetic sector
parameters \\
\hskip 11mm {\tt zm} && $M_Z$, $Z$ boson mass\\
\hskip 11mm {\tt wm} && $M_W$, $W$ boson mass\\
\hskip 11mm {\tt alpha\_{em}} && $\alpha_{em}(M_Z)$, QED coupling at
$M_Z$ scale\\[2mm]
{\tt lam\_fit(alpha\_s)} && Sets $\alpha_s(M_Z)$ and $\Lambda_{QCD}$
for 4-6 flavors at the NNLO level \\
{\tt lam\_fit\_nlo(alpha\_s)} && Sets $\alpha_s(M_Z)$ and
$\Lambda_{QCD}$ for 4-6 flavors at the NLO level \\
\hskip 11mm {\tt alpha\_s} && $\alpha_s(M_Z)$, strong coupling at
$M_Z$ scale\\
\end{tabular}\\[2mm]

\subsubsection{Matter fermion sector}
\label{sec:fmass}

\code{} assumes that neutrinos are massless.  Pole masses of the
charged leptons are initialized in the file\, {\tt mh\_init.f} \,in\,
{\tt block data init\_phys}.  They are stored in the {\tt em} array in
{\tt common/fmass/em(3),um(3),dm(3)} and can be directly modified
there.  Their default values are:\\[2mm]
\begin{tabular}{lp{1mm}l}
Lepton mass && Value \\[2mm]
$m_e$ && {\tt em(1)} = 0.000511  \\
$m_\mu$ && {\tt em(2)} = 0.105659  \\
$m_\tau$ && {\tt em(3)} = 1.777  \\
\end{tabular}\\[2mm]
In the quark sector the most important input parameters are the
running top and bottom masses at a given renormalization scale and the
CKM matrix angles and phase.  They can be set by:\\[2mm]
\begin{tabular}{lp{1mm}p{75mm}}
Routine and arguments && Purpose and MSSM parameters \\[2mm]
{\tt init\_fermion\_sector(tm,tscale,bm,bscale)} && Sets running top
and bottom quark mass \\
\hskip 11mm {\tt tm,tscale} && $m_t(\mu_t)$, running
$\overline{\mathrm{MS}}$ top quark mass\\
\hskip 11mm {\tt bm,bscale} && $m_b(\mu_b)$, running
$\overline{\mathrm{MS}}$ bottom quark mass\\[2mm]
{\tt ckm\_init(s12,s23,s13,delta) } && Initialization of the CKM matrix \\
\hskip 11mm {\tt s12,s23,s13} && $\sin\theta_{12}, \sin\theta_{23},
\sin\theta_{13}$, sines of the CKM angles \\
\hskip 11mm {\tt delta} && $\delta$, the CKM phase in radians\\
\end{tabular}\\[2mm]
The light quark masses are also initialized in the {\tt block data
  init\_phys} of the file {\tt mh\_init.f} and stored in {\tt
  common/fmass\_high/umu(3),uml(3),amuu(3),dmu(3),dml(3),amud(3)}.
The arrays {\tt uml(dml)} contain up(down) quark masses at the scale
{\tt amuu(amud)}, respectively.  Their default values are:\\[2mm]
\begin{tabular}{lp{1mm}lp{1mm}l}
Running quark mass && Mass value && Mass scale \\[2mm]
$m_d(\mu_d)$ && {\tt dml(1)} = 0.007 && {\tt amud(1)} = 2 \\
$m_s(\mu_s)$ && {\tt dml(2)} = 0.11 && {\tt amud(2)} = 2 \\
$m_b(\mu_b)$ && {\tt dml(3)} = 4.17 && {\tt amud(3)} = 4.17 \\
$m_u(\mu_u)$ && {\tt uml(1)} = 0.004 && {\tt amuu(1)} = 2 \\
$m_c(\mu_c)$ && {\tt uml(2)} = 1.279 && {\tt amuu(2)} = 1.279 \\
$m_t(\mu_t)$ && {\tt uml(3)} = 163.5 && {\tt amuu(3)} = 163.5 \\
\end{tabular}\\[2mm]

The variables of the arrays {\tt uml, amuu, dml, amud} can be directly
accessed and modified if necessary.  However, for consistency, after
such modifications the user should call the routine {\tt
  init\_run\_qmass} which calculates running quark masses at the high
$m_t$ scale (stored in {\tt common/fmass\_high/} in the arrays {\tt
  umu,dmu} and in {\tt common/fmass/} in the arrays {\tt um,dm}) for
later use in the running Yukawa couplings and in SUSY loop
calculations.

\subsubsection{Higgs sector}

Following the common convention, we take the Higgs mixing parameter
$\mu$, the CP-odd Higgs boson mass $M_A$, and the ratio of vacuum
expectation values $\tan\beta=v_2/v_1$ as the input parameters.  Other
Higgs sector parameters listed in Section~\ref{sec:parlist} can be
expressed as:
\bea
m_{H_1}^2 &=& \frac{1}{2}(M_A^2 - 2|\mu|^2 - (M_A^2 + M_Z^2)\cos
2\beta)\;,\nonumber\\
m_{H_2}^2 &=& \frac{1}{2}(M_A^2 - 2|\mu|^2 + (M_A^2 + M_Z^2)\cos
2\beta)\;,\nonumber\\
m_{12}^2 &=& - {M_A^2\over 2\sin 2\beta}\;.
\eea
The MSSM Higgs sector at the tree level can be effectively
parametrized in terms of just $M_A$ and $\tan\beta$, but the $\mu$
parameter is necessary for the chargino and neutralino sectors.  Here
it is used to calculate the original Higgs soft mass parameters
$m_{H_1}^2$ and $m_{H_2}^2$ for completeness and future applications;
they currently have no further use.  \\[2mm]
\begin{tabular}{lp{1mm}p{8cm}}
Routine and arguments && Purpose and MSSM parameters \\[2mm]
{\tt init\_higgs\_sector(pm,tb,amu,ierr)} && Higgs sector
initialization \\
\hskip 11mm {\tt pm} && CP-odd Higgs mass $M_A$\\
\hskip 11mm {\tt tb} && Ratio of Higgs VEVs,
$\tan\beta=\frac{v_2}{v_1}$\\
\hskip 11mm {\tt amu} && Higgs mixing parameter $\mu$ (complex)\\
\hskip 11mm {\tt ierr} && output error code: $ierr\neq 0$ if Higgs
sector initialization failed \\[2mm]
{\tt init\_yukawa} && Initialization of the running Yukawa couplings
$Y_l, Y_u, Y_d$ for all generations (at the same scale as the running
quark masses) \\
\end{tabular}

\subsubsection{Supersymmetric fermion sector}

Initialization is done by the routine:\\[2mm]
\begin{tabular}{lp{1mm}p{75mm}}
Routine and arguments && Purpose and MSSM parameters \\[2mm]
{\tt init\_ino\_sector(gm1,gm2,gm3,amu,tb,ierr)} && gaugino sector
initialization \\
\hskip 11mm {\tt gm1,gm2} && $U(1), SU(2)$ gaugino masses (complex) \\
\hskip 11mm {\tt gm3} && $SU(3)$ gaugino mass \\
\hskip 11mm {\tt tb} && $\tan\beta=\frac{v_2}{v_1}$, the ratio of
Higgs VEVs \\
\hskip 11mm {\tt amu} && the Higgs mixing parameter $\mu$ (complex) \\
\hskip 11mm {\tt ierr} && output warning code: $ierr\neq 0$ if a
char\-gi\-no or a neutralino is lighter than $M_Z/2$ \\[2mm]
\end{tabular}\\[2mm]
If one sets $M_1=0$ in the call to {\tt init\_ino\_sector} then the
GUT-derived relation $M_1 = \frac{5}{3}\tan^2\theta_W M_2$ is used in
the gaugino mass calculations.

\subsubsection{Sfermion sector}  
\label{sec:sferinit}

This is the most complicated MSSM sector; it contains a large number
of free parameters.  \code{} supplies two subroutines for the sfermion
parameters initialization, {\tt init\_slepton\_sector} and {\tt
  init\_squark\_sector}.  They accept as input only dimensionless mass
insertions and dimensionless diagonal trilinear soft mixing terms,
expanded in \code{} to entries of the soft mass matrices as defined by
eqs.~(\ref{eq:midef},\ref{eq:lrdef},\ref{eq:lrmidef}) (this is only a
particular choice of parametrization and does not lead to any loss of
generality). The sfermion initialization routines have the following
arguments:\\[2mm]
\begin{tabular}{lp{1mm}p{132mm}}
\multicolumn{3}{l}{\tt subroutine
  init\_slepton\_sector(sll,slr,asl,ierr,slmi\_l,slmi\_r)}\\[2mm]
Argument && MSSM parameters \\[2mm]
{\tt sll} && Array of the diagonal left slepton masses
$(m_L^2)_{II}={\tt sll(I)}^2$, $I=1\ldots 3$\\
{\tt slr} && Array of the diagonal right slepton masses
$(m_E^2)_{II}={\tt slr(I)}^2$, $I=1\ldots 3$\\
{\tt asl} && Array of the dimensionless diagonal slepton trilinear
mixing terms $a_l^I={\tt asl(I)}$, $I=1\ldots 3$ (complex
parameters).\\
{\tt ierr} && output error code: $ierr\neq0$ if slepton sector
initialization failed \\
{\tt slmi\_l} && Array of the off-diagonal left slepton mass
insertions $\delta^{12}_{L} = {\tt slmi\_l(1)}$, $\delta^{23}_{L} =
{\tt slmi\_l(2)}$, $\delta^{13}_{L} = {\tt slmi\_l(3)}$ (complex
parameters); remaining LL mass insertions are initialized via
hermitian conjugation\\
{\tt slmi\_r} && Array of the off-diagonal right slepton mass
insertions $\delta^{12}_{E} = {\tt slmi\_r(1)}$, $\delta^{23}_{E} =
{\tt slmi\_r(2)}$, $\delta^{13}_{E} = {\tt slmi\_r(3)}$ (complex
parameters); remaining RR mass insertions are initialized via
hermitian conjugation\\
{\tt slmi\_lr} && Matrix with off-diagonal slepton trilinear LR mass
insertions $\delta^{IJ}_{LLR} = {\tt slmi\_lr(I,J)}$, $I,J=1\ldots 3$
(complex parameters)\\
\end{tabular}

\begin{tabular}{lp{1mm}p{132mm}}
\multicolumn{3}{l}{\tt subroutine
  init\_squark\_sector(sql,squ,sqd,asu,asd,ierr, }\\
&&{\tt sqmi\_l,sumi\_r,sdmi\_r,sumi\_lr,sdmi\_lr) }\\[2mm]
Argument && MSSM parameters \\[2mm]
{\tt sql} && Array of the diagonal left squark masses
$(m_Q^2)_{II}={\tt sql(I)}^2$, $I=1\ldots 3$\\
{\tt squ} && Array of the diagonal right up-squark masses
$(m_U^2)_{II}={\tt squ(I)}^2$, $I=1\ldots 3$\\
{\tt sqd} && Array of the diagonal right down-squark masses
$(m_D^2)_{II}={\tt sqd(I)}^2$, $I=1\ldots 3$\\
{\tt asu} && Array of the dimensionless diagonal soft LR up-squark
mixing terms $a_u^I={\tt asu(I)}$, $I=1\ldots 3$ (complex
parameters)\\
{\tt asd} && Array of the dimensionless diagonal soft LR down-squark
mixing terms $a_d^I={\tt asd(I)}$, $I=1\ldots 3$ (complex
parameters)\\
{\tt ierr} && output error code: $ierr\neq 0$ if squark sector
initialization failed \\
{\tt sqmi\_l} && Array of the off-diagonal left squark mass insertions
$\delta^{12}_{Q} = {\tt sqmi\_l(1)}$, $\delta^{23}_{Q} = {\tt
  sqmi\_l(2)}$, $\delta^{13}_{Q} = {\tt sqmi\_l(3)}$ (complex
parameters); remaining QLL mass insertions are initialized via
hermitian conjugation\\
{\tt sumi\_r} && Array of the off-diagonal right up-squark mass
insertions $\delta^{12}_{U} = {\tt sumi\_r(1)}$, $\delta^{23}_{U} =
{\tt sumi\_r(2)}$, $\delta^{13}_{U} = {\tt sumi\_r(3)}$ (complex
parameters); remaining URR mass insertions are initialized via
hermitian conjugation\\
{\tt sdmi\_r} && Array of the off-diagonal right down-squark mass
insertions $\delta^{12}_{D} = {\tt sdmi\_r(1)}$, $\delta^{23}_{D} =
{\tt sdmi\_r(2)}$, $\delta^{13}_{D} = {\tt sdmi\_r(3)}$ (complex
parameters); remaining DRR mass insertions are initialized via
hermitian conjugation\\
{\tt sumi\_lr} && Matrix with off-diagonal up-squark trilinear LR mass
insertions $\delta^{IJ}_{ULR} = {\tt sumi\_lr(I,J)}$, $I,J=1\ldots 3$
(complex parameters)\\
{\tt sdmi\_lr} && Matrix with off-diagonal down-squark trilinear LR
mass insertions $\delta^{IJ}_{DLR} = {\tt sdmi\_lr(I,J)}$,
$I,J=1\ldots 3$ (complex parameters)\\
\end{tabular}\\[2mm]

If necessary, experienced \code{} users can directly modify the soft
breaking sfermion parameters stored in common blocks {\tt /msoft/} and
{\tt /soft/} (see Table~\ref{tab:mssmpar} in
Section~\ref{sec:commons}). One must subsequently call the routines
{\tt sldiag, sqdiag} (see file {\tt mh\_diag.f}) to recalculate the
tree-level sfermion masses and mixing matrices.  This may, however,
require a deeper understanding of the \code{} initialization sequence
and its data structure.

\subsection{Tree-level physical masses and mixing angles}
\label{sec:commons}

After performing the full initialization sequence in \code{}, all the
MSSM Lagrangian parameters listed in Section~\ref{sec:parlist},
physical tree-level particle masses (with the exception of the
running quark masses), and mixing matrices are calculated and stored
in common blocks.  If necessary, they can be directly accessed and
modified.  Note, however, that after any modifications of the
Lagrangian parameters, relevant procedures calculating physical masses
and mixing angles have to called again.  In Table~\ref{tab:mssmpar} we
list the important blocks storing MSSM parameters.  Common blocks
containing masses and mixing angles are listed in
Table~\ref{tab:eigenmass}.

\begin{table}[htbp]
\begin{center}
\begin{tabular}{lp{1mm}p{85mm}}
Common block and variables && Lagrangian parameters \\[2mm]
\multicolumn{3}{l}{\tt
  common/vpar/st,ct,st2,ct2,sct,sct2,e,e2,alpha,wm,wm2,zm,zm2,pi,sq2}
\\
\hskip 9mm {\tt st,ct,st2,ct2,sct,sct2} && Weinberg angle functions,
respectively $s_W$, $c_W$, $s_W^2$, $c_W^2$, $s_Wc_W$, $s_W^2 c_W^2$\\
\hskip 9mm {\tt e,e2,alpha} && electric charge powers at $M_Z$ scale:
$e$, $e^2$, $\alpha_{em}$ \\
\hskip 9mm {\tt wm,wm2,zm,zm2} && gauge boson masses: $M_W$, $M_W^2$,
$M_Z$, $M_Z^2$\\
\hskip 9mm {\tt pi,sq2} && numerical constants, $\pi$ and $\sqrt{2}$
\\[2mm]
\multicolumn{3}{l}{\tt common/hpar/hm1,hm2,hm12,hmu} \\
\hskip 9mm {\tt hm1,hm2} && soft Higgs masses $m_{H_1}^2,
m_{H_2}^2$\\
\hskip 9mm {\tt hm12} && soft Higgs mixing parameter $m_{12}^2$\\
\hskip 9mm {\tt hmu} && Higgs mixing parameter $\mu$ (complex)\\[2mm]
\multicolumn{3}{l}{\tt common/vev/v1,v2} \\
\hskip 9mm {\tt v1,v2} && Higgs vacuum expectation values $v_1,
v_2$\\[2mm]
\multicolumn{3}{l}{\tt common/yukawa/yl(3),yu(3),yd(3)} \\
\hskip 9mm {\tt yl(3)} && charged lepton Yukawa couplings $Y_e$,
$Y_\mu$, $Y_\tau$\\
\hskip 9mm {\tt yu(3)} && Running $\overline{\mathrm{MS}}$ up-quark
Yukawa couplings at $m_t$ scale: $Y_u$, $Y_c$, $Y_t$ \\
\hskip 9mm {\tt yd(3)} && Running $\overline{\mathrm{MS}}$ down-quark
Yukawa couplings at $m_t$ scale: $Y_u,Y_c,Y_t$ \\[2mm]
\multicolumn{3}{l}{\tt common/gmass/gm3,gm2,gm1} \\
\hskip 9mm {\tt gm1,gm2} && $U(1),SU(2)$ gaugino masses $M_1,M_2$
(complex)\\
\hskip 9mm {\tt gm3} && $SU(3)$ gaugino mass $M_3$\\[2mm]
\multicolumn{3}{l}{\tt
  common/msoft/lms(3,3),rms(3,3),ums(3,3),dms(3,3),qms(3,3)}\\
\hskip 9mm {\tt lms(3,3),rms(3,3)} && hermitian slepton soft mass
matrices $m_L^2$, $m_E^2$ (complex)\\
\hskip 9mm {\tt ums(3,3),dms(3,3),qms(3,3)} && hermitian squark soft
mass matrices $m_U^2$, $m_D^2$, $m_Q^2$ (complex)\\[2mm]
\multicolumn{3}{l}{\tt
  common/soft/ls(3,3),ks(3,3),ds(3,3),es(3,3),us(3,3),ws(3,3)}\\
\hskip 9mm {\tt ls(3,3),ds(3,3),us(3,3)} && trilinear soft LR mixing
matrices $A_l$, $A_d$, $A_u$ (complex) \\
\hskip 9mm {\tt ks(3,3),es(3,3),ws(3,3)} && trilinear
``non-holomorphic'' soft mixing matrices $A'_l$, $A'_d$, $A'_u$
(complex) \\
\end{tabular}
\end{center}
\caption{Common blocks storing the MSSM Lagrangian parameters.  We
  omit flavor indices in the fermion and sfermion sectors.}
\label{tab:mssmpar}
\end{table}

\begin{table}[htbp]
\begin{center}
\begin{tabular}{lp{1mm}p{85mm}}
Common block and variables && Masses and mixing matrices \\[2mm]
\multicolumn{3}{l}{\tt common/fmass/em(3),um(3),dm(3)} \\
\hskip 9mm {\tt em(3)} && Charged lepton pole masses $m_e,m_\mu,m_\tau$ \\
\hskip 9mm {\tt um(3)} && Running $\overline{\mathrm{MS}}$ up-quark
masses at the $m_t$ scale: $m_u,m_c,m_t$ \\
\hskip 9mm {\tt dm(3)} && Running $\overline{\mathrm{MS}}$ down-quark
masses at the $m_t$ scale: $m_u,m_c,m_t$ \\[2mm]
\multicolumn{3}{l}{\tt common/hmass/cm(2),rm(2),pm(2),zr(2,2),zh(2,2)} \\
\hskip 9mm {\tt rm(2)} && neutral CP-even Higgs masses ${\tt rm(1)} =
M_H$, ${\tt rm(2)} = M_h$ \\
\hskip 9mm {\tt pm(2)} && neutral CP-odd Higgs mass {\tt pm(1)} and
Goldstone mass {\tt pm(2)} \\
\hskip 9mm {\tt cm(2)} && charged Higgs mass {\tt cm(1)} and charged
Goldstone mass {\tt cm(2)} \\
\hskip 9mm {\tt zr(2,2)} && CP-even Higgs mixing matrix $Z_R$ \\
\hskip 9mm {\tt zh(2,2)} && CP-odd and charged Higgs mixing matrix
$Z_H$ \\[2mm]
\multicolumn{3}{l}{\tt common/charg/fcm(2),zpos(2,2),zneg(2,2)} \\
\hskip 9mm {\tt fcm(2)} && chargino masses $M_{\chi^+_i}$, $i=1,2$ \\
\hskip 9mm {\tt zpos(2,2),zneg(2,2)} && chargino mixing matrices $Z_+,
Z_-$ (complex) \\[2mm]
\multicolumn{3}{l}{\tt common/neut/fnm(4),zn(4,4)}\\
\hskip 9mm {\tt fnm(4)} && neutralino masses $M_{\chi^0_i}$,
$i=1\ldots 4$ \\
\hskip 9mm {\tt zn(4,4)} && neutralino mixing matrix $Z_N$ (complex)
\\[2mm]
\multicolumn{3}{l}{\tt common/slmass/vm(3),slm(6),zv(3,3),zl(6,6)} \\
\hskip 9mm {\tt vm(3)} && sneutrino masses $M_{\tilde\nu_I}, I=1\ldots
3$ \\
\hskip 9mm {\tt slm(6)} && charged slepton masses $M_{L_i}, i=1\ldots
6$ \\
\hskip 9mm {\tt zv(3,3)} && sneutrino mixing matrix $Z_{\tilde\nu}$
(complex)\\
\hskip 9mm {\tt zl(6,6)} && charged slepton mixing matrix $Z_L$
(complex) \\[2mm]
\multicolumn{3}{l}{\tt common/sqmass/sum(6),sdm(6),zu(6,6),zd(6,6)} \\
\hskip 9mm {\tt sum(6)} && up-squark masses $M_{U_i}, i=1\ldots 6$ \\
\hskip 9mm {\tt sdm(6)} && down-squark masses $M_{D_i}, i=1\ldots 6$
\\
\hskip 9mm {\tt zu(6,6)} && up-squark mixing matrix $Z_U$ (complex) \\
\hskip 9mm {\tt zd(6,6)} && down-squark mixing matrix $Z_D$ (complex)
\\
\end{tabular}
\end{center}
\caption{Common blocks storing physical particle masses and mixing
  matrices.}
\label{tab:eigenmass}
\end{table}

All parameters, tree-level masses, and mixing angles can be printed
for test purposes, e.g.  by calling the subroutines {\tt
  print\_MSSM\_par} and {\tt print\_MSSM\_masses}.

\section{List of processes}
\label{sec:proc}

In this section we list the set observables whose computation is fully
implemented in \code{} v1.0.  For all of them, \code{} takes into
account one-loop supersymmetric contributions.  QCD corrections and
hadronic matrix elements are extracted from the papers of various
authors, mostly from analyses done in the Standard Model.  They are
assumed to work reasonably well also in the MSSM since supersymmetric
strong corrections from gluino and squarks are suppressed by large
masses of these particles.

In most cases, QCD and hadronic corrections are known at the level of
few to tens \%, while variations of supersymmetric flavor and CP
violating parameters can change observables by orders-of-magnitude.
Thus, as long as the MSSM parameters are not measured very precisely,
the current implementation of strong corrections is sufficient for
analyses performed in the framework of the general MSSM.

Calculations of the hadronic matrix elements are particularly
difficult as they have to be performed, at least partially, in the
regime of strongly coupled QCD.  Results of such calculations can
differ significantly depending on the methods used and thus carry
significant theoretical uncertainties.  Therefore, in \code,
quantities which requires hadronic matrix element estimates and other
QCD related quantities are treated as external parameters.  They are
initialized to the default values listed below for each observable and
can be directly modified by users by changing the relevant variables
in the common blocks where they are stored.  Currently most of the
hadronic (and related) input parameters used in \code{} are taken from
the Table~3 of Ref.~\cite{AB}.

\subsection{Electric Dipole Moments of charged leptons}
\label{sec:edml}

Lepton EDMs are defined as the coefficient $d_{l^I}$ in the effective
Hamiltonian for the flavor-diagonal lepton-lepton-photon interaction:
\bea \label{eq:edm}
{\cal H}_e = {i d_{l^I}\over 2}\bar l^I \sigma_{\mu\nu} \gamma_5 l^I
F^{\mu\nu}\;,
\eea
where $I=1,2,3$ is the generation index of the lepton as usual.  In
\code{} lepton EDM is calculated by:\\[2mm]
\begin{tabular}{lp{5mm}p{95mm}}
Routine: && {\tt double precision function edm\_l(I)} \\
Input: && $I=1,2,3$ for $e,\mu,\tau$ respectively\\
Output: && EDM for the charged lepton specified by $I$\\
QCD related factors: && none, QCD corrections are small and not
included\\
Details of calculations: && Ref.~\cite{POROSA}\\
\end{tabular}

\subsection{Neutron Electric Dipole Moment}
\label{sec:edmn}

The neutron EDM can be approximated by the sum of the electric dipole
moments of the constituent $d$ and $u$ quarks plus contributions of
the chromoelectric dipole moments (CDM) of quarks and gluons.  The
EDMs of the individual quarks are defined analogously to
eq.~(\ref{eq:edm}).  The CDM $c_q$ of quark $q$ is defined as:
\bea
{\cal H}_c = - \frac{i c_q}{2} \bar{q} \sigma_{\mu\nu} \gamma_5 T^a q
G^{\mu\nu a}.
\label{eq:cdmdef}
\eea
The gluonic dipole moment $c_g$ is defined as:
\bea
{\cal H}_g = - \frac{c_g}{6} f_{abc} G^a_{\mu\rho} G^{b\rho}_{\nu}
G^c_{\lambda\sigma}\epsilon^{\mu\nu\lambda\sigma}.
\label{eq:gdmdef}
\eea
The exact calculation of the neutron EDM requires knowledge of its
hadronic wave function.  \code{} uses the ``na\"ive'' chiral quark
model approximation~\cite{QUARKMODEL}:
\bea
E_n = \frac{\eta_e}{3}(4d_d - d_u) + \frac{e\eta_c}{4\pi}(4c_d - c_u)
+\frac{e\eta_g\Lambda_X}{4\pi}c_g
\label{eq:fullneut}
\eea
where $\eta_i$ and $\Lambda_X$ are the QCD correction
factors~\cite{ETAQCD} and the chiral symmetry breaking
scale~\cite{QUARKMODEL}, respectively.  Various models give
significantly different $\eta_i$ factors.  As a result, even the sign
of the neutron EDM is not certain.  Thus the \code{} result should be
treated as an order of magnitude estimate only.  The calculations are
performed by calling\\[2mm]
\begin{tabular}{lp{5mm}p{95mm}}
Routine && {\tt double precision function edm\_n()} \\
Input && none\\
Output && neutron EDM\\
QCD related factors: \\
\multicolumn{3}{l}
{\tt common/edm\_qcd/eta\_e,eta\_c,eta\_g,alamx} \\
\hskip 9mm $\eta_{e}$ && \hskip 9mm ${\tt eta\_e} = 1.53 $\\
\hskip 9mm $\eta_{g}$ && \hskip 9mm ${\tt eta\_c} = 3.4 $\\
\hskip 9mm $\eta_{g}$ && \hskip 9mm ${\tt eta\_g} = 3.4 $\\
\hskip 9mm $\Lambda_{X}$ && \hskip 9mm ${\tt alamx} = 1.18 $\\
Details of calculations: && Ref.~\cite{POROSA}\\
\end{tabular}

\subsection{$K^0_L\ra \pi^0 \bar\nu \nu$ and $K^+\ra \pi^+ \bar\nu \nu$
decay rates}
\label{sec:kpivv}

The relevant part of the effective Hamiltonian generated by the top
quark and SUSY particle exchanges can be written as
\begin{equation}\label{eq:kpivvham} 
{\cal H}_{\rm eff}={G_{\rm F} \over{\sqrt 2}}{\alpha\over 2\pi
  \sin^2\theta_{\rm w}} \sum_{l=e,\mu,\tau} \left[X_L (\bar
  sd)_{V-A}(\bar\nu_l\nu_l)_{V-A}+ X_R (\bar
  sd)_{V+A}(\bar\nu_l\nu_l)_{V-A}\right].
\end{equation}
The branching ratios for the $K\ra \pi \nu \bar \nu$ decays are then
given by
\bea
\label{bkpnZ}
{Br}(K^+\ra \pi^+ \bar\nu \nu) = \kappa_+ \left[ \left({ \imag
      (X_L+X_R) \over \lambda^5} \right)^2 + \left( {\real
      (K^{\star}_{cs}K_{cd}) \over \lambda} P_c + {\real (X_L + X_R)
      \over\lambda^5} \right)^2 \right]
\eea
\bea
\label{bklpnZ}
\mathrm{Br}(K^0_L\ra \pi^0 \bar\nu \nu)=\kappa_L \left({\imag (X_L +
    X_R) \over \lambda^5}\right)^2
\eea
where $\kappa$~\cite{Mescia:2007kn}, $\lambda$ (one of the Wolfenstein
parameters~\cite{WO}), and the NLO charm quark contribution
$P_c$~\cite{BB98,BB3,GORBAHN} can be modified by \code{} users (note
that $\kappa$ and $P_c$ depend on $V_{us}$, $m_c$ and $\alpha_s$)
%
%
%
%
%
Branching ratio calculations are performed by calling \\[2mm]
\begin{tabular}{lp{5mm}p{95mm}}
Routine && {\tt subroutine k\_pivv(br\_k0,br\_kp)} \\
Input && none\\
Output && ${\tt br\_k0} = Br(K^0_L\ra \pi^0 \bar\nu \nu)$ \\
       && ${\tt br\_kp} = Br(K^+\ra \pi^+ \bar\nu \nu)$\\
QCD related factors \\
\multicolumn{3}{l}
{\tt common/kpivv/ak0,del\_ak0,akp,del\_akp,pc,del\_pc,alam} \\
\hskip 9mm $\kappa_L \pm \Delta\kappa_L$ && \hskip 9mm ${\tt
  ak0}=2.231\cdot 10^{-10}$, ${\tt del\_ak0}=0.013\cdot 10^{-10}$ \\
\hskip 9mm $\kappa_+ \pm \Delta\kappa_+$ && \hskip 9mm ${\tt
  akp}=5.173\cdot 10^{-11}$, ${\tt del\_akp}=0.025\cdot 10^{-11}$\\
\hskip 9mm $P_c \pm \Delta P_c$ && \hskip 9mm ${\tt pc}=0.41$, ${\tt
  del\_pc}=0.03$\\
\hskip 9mm $\lambda$ && \hskip 9mm ${\tt alam}=0.225$\\
Details of calculations: && Ref.~\cite{BEJR}\\
\end{tabular}

\subsection{$B_d^0\ra l^{I+} l^{J-}$ and $B_s^0\ra l^{I+} l^{J-}$
  decay rates}
\label{sec:bll}

The general expression for these branching ratios are rather
complicated and can be found in~\cite{DRT}.  For most users it is
sufficient to know that, in addition to the MSSM parameters, the
dilepton $B$ decays depend on the $B$ meson masses and the hadronic
matrix elements of the down quark vector and scalar currents:
\bea
\bra{0}\overline{b}\gamma_{\mu} P_{L(R)} s \ket{B_{s(d)}(p)} \ &=&
\ -(+) \frac{i}{2} p_{\mu} f_{B_{s(d)}} \;, \label{np1} \\[3mm]
\bra{0}\overline{b} P_{L(R)} s \ket{B_{s(d)}(p)} \ &=& \ +(-)
\frac{i}{2} \, \frac{M_{B_{s(d)}}^{2} f_{B_{s}}}{m_{b}+m_{s(d)}}
\;,  \label{np2}
\eea
where $p_{\mu}$ is the momentum of the decaying $B_{s(d)}$-meson of
mass $M_{B_{s(d)}}$.  The $B_d^0\ra l^{I+} l^{J-}$ and $B_s^0\ra
l^{I+} l^{J-}$ decay branching ratios are calculated by:\\[2mm]
\begin{tabular}{lp{5mm}p{105mm}}
Routine && {\tt double precision function b\_ll(K,L,I,J)} \\
Input && $I,J=1,2,3$ - outgoing leptons generation indices \\
&& $K,L$ - generation indices of the valence quarks of the $B^0$
meson: setting $(K,L) = (3,1), (1,3), (3,2)$ and $(2,3)$ chooses
respectively $B_d^0$, $\bar B_d^0$, $B_s^0$ and $\bar B_s^0$ decay \\
Output && Branching ratios of the decay defined by $K,L,I,J$\\
QCD related factors \\
\multicolumn{3}{l}
{\tt
  common/meson\_data/dmk,amk,epsk,fk,dmd,amd,fd,amb(2),dmb(2),gam\_b(2),fb(2)}
\\
\hskip 9mm $M_{B_d}$ && \hskip 9mm ${\tt amb(1)}=5.2794$ \\
\hskip 9mm $M_{B_s}$ && \hskip 9mm ${\tt amb(2)}=5.368$ \\
\hskip 9mm $f_{B_d}$ && \hskip 9mm ${\tt fb(1)}=0.2$ \\
\hskip 9mm $f_{B_s}$ && \hskip 9mm ${\tt fb(2)}=0.245$ \\
Details of calculations: && Ref.~\cite{DRT}\\
\end{tabular}

\subsection{$\bar K^0 K^0$ meson mixing parameters}
\label{sec:kkmix}

\code{} calculates two parameters measuring the amount of CP-violation
in neutral $K$ meson oscillations: $\varepsilon_K$ and the $\bar
K^0-K^0$ mass difference $\Delta M_K$.
\bea
\Delta M_K= 2 \real\langle \bar K^0| H^{\Delta S=2}_{\rm
  eff}|K^0\rangle~,
\eea
\bea
\varepsilon_K=\frac{\exp(i\pi/4)}{\sqrt{2}\Delta M_K} \imag\langle
\bar K^0| H^{\Delta S=2}_{\rm eff}|K^0\rangle~.
\eea
QCD dependent corrections are known with reasonable accuracy for the
$\varepsilon_K$ parameter.  The long distance contributions to $\Delta
M_K$ are large and difficult to control.  Thus the result given by
\code{} for $\Delta M_K$ should be treated as an order of magnitude
estimate only.

Apart from the MSSM parameters, the calculation of the $\bar K^0 K^0$
meson mixing requires knowledge of the meson masses and of the
hadronic matrix elements of the following set of four-quark operators:
\bea 
Q_1^{\rm VLL} &=& (\bar{q}^I_{\alpha} \gamma_{\mu}    P_L q^J_{\alpha})
                  (\bar{q}^I_{\beta} \gamma^{\mu}    P_L q^J_{\beta}),
\nonumber\\ 
Q_1^{\rm LR} &=&  (\bar{q}^I_{\alpha} \gamma_{\mu}    P_Lq^J_{\alpha})
                  (\bar{q}^I_{\beta} \gamma^{\mu}    P_R q^J_{\beta}),
\nonumber\\
Q_2^{\rm LR} &=&  (\bar{q}^I_{\alpha}                 P_Lq^J_{\alpha})
                  (\bar{q}^I_{\beta}                 P_R q^J_{\beta}),
\nonumber\\
Q_1^{\rm SLL} &=& (\bar{q}^I_{\alpha}                 P_Lq^J_{\alpha})
                  (\bar{q}^I_{\beta}                 P_L q^J_{\beta}),
\nonumber\\
Q_2^{\rm SLL} &=& (\bar{q}^I_{\alpha} \sigma_{\mu\nu} P_Lq^J_{\alpha})
                  (\bar{q}^i_{\beta} \sigma^{\mu\nu} P_L q^J_{\beta})
\label{eq:4qbase}
\eea
where $\alpha, \beta$ are color indices, for the $\bar{K}^0 K^0$
mixing one should choose flavor indices $I=2$ and $J=1$.  The matrix
elements can be written as:
\bea
\label{eq:4qmatel}
\langle \bar K^0 | Q_1^{\rm VLL}(\mu) | K^0 \rangle &=& \frac{1}{3}
M_K F_K^2 B_1^{\rm VLL} (\mu) ,\nonumber\\
\langle \bar K^0 | Q_1^{\rm LR}(\mu) | K^0 \rangle &=& -\frac{1}{6}
\left( \frac{M_K}{m_s(\mu) + m_d(\mu)} \right)^2 M_K F_K^2 B_1^{\rm
  LR} (\mu) ,\nonumber\\
\langle \bar K^0 | Q_2^{\rm LR}(\mu) | K^0 \rangle &=& \frac{1}{4}
\left( \frac{M_K}{m_s(\mu) + m_d(\mu)} \right)^2 M_K F_K^2 B_2^{\rm
  LR} (\mu) ,\nonumber\\
\langle \bar K^0 | Q_1^{\rm SLL}(\mu) | K^0 \rangle &=& -\frac{5}{24}
\left( \frac{M_K}{m_s(\mu) + m_d(\mu)} \right)^2 M_K F_K^2 B_1^{\rm
  SLL} (\mu) ,\nonumber\\
\langle \bar K^0 | Q_2^{\rm SLL}(\mu) | K^0 \rangle &=& -\frac{1}{2}
\left( \frac{M_K}{m_s(\mu) + m_d(\mu)} \right)^2 M_K F_K^2 B_2^{\rm
  SLL} (\mu),
\eea
where $F_K$ is the $K$-meson decay constant.  By default, \code{} uses
the $B_i^X$ values at the scale $\mu=2$ GeV given in~\cite{BJU} using
the NDR renormalization scheme (quark masses at the scale 2 GeV are
stored in {\tt common/fmass\_high/}, see Section~\ref{sec:fmass}).

In addition to the hadronic matrix elements, QCD corrections depend
also on the ``$\eta$'' factors describing the evolution of the
relevant Wilson coefficients from the high to low energy scale.  These
factors are automatically calculated at NLO by \code{}.  For the SM
contribution to the Wilson coefficient of the $Q^{\mathrm{VLL}}$
operator a separate careful calculation of the evolution factors has
been performed~\cite{BJW,HN}.  Therefore \code{} treats this
contribution separately, setting $B_{SM}^{\rm VLL}$ and the
$\eta_{SM}$ factor to default values given in~\cite{Buras9806471}
(see~\cite{BJU} for a very detailed discussion of the structure of the
QCD corrections in $\bar{B}^0B^0$ and $\bar{K}^0K^0$ systems,
including their renormalization scheme dependence and calculations of
the evolution factors ``$\eta$'' implemented in \code).

The kaon mass difference $\Delta M_K$ and the $\varepsilon_K$
parameter measuring the amount of CP violation in $\bar K^0 K^0$
mixing are calculated by\\[2mm]
\begin{tabular}{lp{5mm}p{95mm}}
Routine && {\tt subroutine dd\_kaon(eps\_k,delta\_mk)} \\
Input && none\\
Output && ${\tt eps\_k}=\varepsilon_K$ parameter \\
 && ${\tt delta\_mk }=\Delta M_K$ mass difference\\
QCD related factors: \\
\multicolumn{3}{l}
{\tt
  common/meson\_data/dmk,amk,epsk,fk,dmd,amd,fd,amb(2),dmb(2),gam\_b(2),fb(2)}
\\
\hskip 9mm $M_K$ && \hskip 9mm ${\tt amk}=0.497672$ \\
\hskip 9mm Measured $\Delta M_K^{exp}$ && \hskip 9mm ${\tt dmk}=
3.49\cdot 10^{-15}$ \\
\hskip 9mm Measured $\varepsilon_K^{exp}$ && \hskip 9mm ${\tt
  epsk}=2.26\cdot 10^{-3}$ \\
\hskip 9mm $f_K$ && \hskip 9mm ${\tt fk}=0.1598$ \\
\multicolumn{3}{l}
{\tt common/bx\_4q/bk(5),bd(5),bb(2,5),amu\_k,amu\_d,amu\_b}\\
\hskip 9mm $B_1^{\rm VLL} (\mu_K)$ && \hskip 9mm ${\tt bk(1)}=0.61$\\
\hskip 9mm $B_1^{\rm SLL} (\mu_K)$ && \hskip 9mm ${\tt bk(2)}=0.76$\\
\hskip 9mm $B_2^{\rm SLL} (\mu_K)$ && \hskip 9mm ${\tt bk(3)}=0.51$\\
\hskip 9mm $B_1^{\rm LR} (\mu_K)$ && \hskip 9mm ${\tt bk(4)}=0.96$\\
\hskip 9mm $B_2^{\rm LR} (\mu_K)$ && \hskip 9mm ${\tt bk(5)}=1.30$\\
\hskip 9mm Renormalization scale $\mu_K$ && \hskip 9mm ${\tt
  amu\_k}=2$\\
\multicolumn{3}{l}
{\tt
  common/sm\_4q/eta\_cc,eta\_ct,eta\_tt,eta\_b,bk\_sm,bd\_sm,bb\_sm(2)}\\
\hskip 9mm $B_{SM}^{\rm VLL}$ && \hskip 9mm ${\tt bk\_sm}=0.724$\\
\hskip 9mm $\eta_{cc}$ && \hskip 9mm ${\tt eta\_cc}=1.44$\\
\hskip 9mm $\eta_{ct}$ && \hskip 9mm ${\tt eta\_ct}=0.47$\\
\hskip 9mm $\eta_{tt}$ && \hskip 9mm ${\tt eta\_tt}=0.57$\\
Details of calculations: && Ref.~\cite{BJU,BCRS}\\
\end{tabular}

\subsection{$\bar D^0 D^0$ meson mass difference}
\label{sec:ddmix}

Calculations of the mass difference $\Delta m_D$ of the neutral $D$
mesons have large theoretical uncertainties due to unknown
long-distance strong corrections.  Thus, as in the case of $\Delta
m_K$, the \code{} result for $\Delta m_D$ should be treated as an
order of magnitude estimate only.

The structure of strong corrections is analogous to those in the $K$
meson system.  However, in this case hadronic matrix elements and QCD
evolution calculations available in the literature are much less
refined.  \code{} uses the NLO evolution for the ``$\eta$'' factors
and sets, by default, all the relevant hadronic matrix elements
$B_i=1$, i.e.  it uses the ``vacuum saturation'' approximation (this
can be changed easily when new results become available).\\[2mm]
\begin{tabular}{lp{5mm}p{95mm}}
Routine && {\tt subroutine uu\_bmeson(delta\_md)} \\
Input && none\\ 
Output && ${\tt delta\_md }=\Delta M_D$ mass difference\\
QCD related factors: \\
\multicolumn{3}{l}
{\tt
  common/meson\_data/dmk,amk,epsk,fk,dmd,amd,fd,amb(2),dmb(2),gam\_b(2),fb(2)}
\\
\hskip 9mm $M_D$ && \hskip 9mm ${\tt amd}=1.8645$ \\
\hskip 9mm Measured $\Delta M_D^{exp}$ && \hskip 9mm ${\tt dmd}=
4.61\cdot 10^{-14}$ \\
\hskip 9mm $f_D$ && \hskip 9mm ${\tt fd}=0.165$ \\
\multicolumn{3}{l}
{\tt common/bx\_4q/bk(5),bd(5),bb(2,5),amu\_k,amu\_d,amu\_b}\\
\hskip 9mm $B_1^{\rm VLL} (\mu_D)$ && \hskip 9mm ${\tt bd(1)}=1$\\
\hskip 9mm $B_1^{\rm SLL} (\mu_D)$ && \hskip 9mm ${\tt bd(2)}=1$\\
\hskip 9mm $B_2^{\rm SLL} (\mu_D)$ && \hskip 9mm ${\tt bd(3)}=1$\\
\hskip 9mm $B_1^{\rm LR} (\mu_D)$ && \hskip 9mm ${\tt bd(4)}=1$\\
\hskip 9mm $B_2^{\rm LR} (\mu_D)$ && \hskip 9mm ${\tt bd(5)}=1$\\
\hskip 9mm Renormalization scale $\mu_D$ && \hskip 9mm ${\tt
  amu\_d}=2$\\
\multicolumn{3}{l}
{\tt
  common/sm\_4q/eta\_cc,eta\_ct,eta\_tt,eta\_b,bk\_sm,bd\_sm,bb\_sm(2)}\\
\hskip 9mm $B_{SM}^{\rm VLL}$ && \hskip 9mm ${\tt bd\_sm}=1$\\
Details of calculations: && Performed by authors, unpublished \\
\end{tabular}

\subsection{$\bar B_d^0 B_d^0$ and $\bar B_s^0 B_s^0$ mass differences}
\label{sec:bbmix}

Mixing and CP violation phenomena are also observed in the neutral $B$
meson systems.  In particular, the mass differences in the $\bar B_d^0
B_d^0$ and $\bar B_s^0 B_s^0$ oscillations have been measured,
\bea
\Delta M_{B_{d(s)}} = 2 |\langle \bar B^0_{d(s)}| H^{\Delta B=2}_{\rm
  eff}|B^0_{d(s)}\rangle|~.
\label{eq:bbmix}
\eea
In addition to the MSSM parameters, theoretical calculations of
$\Delta m_{B_d}$ and $\Delta m_{B_s}$ depend, as for $K$ and $D$
oscillations, on the relevant hadronic matrix elements and QCD
evolution factors.  The formulae for $\bar{B}^0B^0$ mixing can be
obtained by making the obvious replacements in the formulae presented
in Section~\ref{sec:kkmix}.  Currently \code{} uses the same set of
$B_i$ factors for both the $B_d$ and $B_s$ sectors, but it leaves the
possibility to distinguish between them in future, if necessary.  For
this one needs to independently initialize the arrays {\tt bb(1,i)}
($B_d$ meson hadronic matrix elements) and {\tt bb(2,i)} ($B_s$ meson
hadronic matrix elements) stored in {\tt common/bx\_4q/}.

The values of the $B$ meson masses and coupling constants are the same
as those listed in Section~\ref{sec:bll}.  \code{} calculates the mass
differences $\Delta M_{B_{d(s)}}$ as defined by eq.~(\ref{eq:bbmix}):
\\[2mm]
\begin{tabular}{lp{5mm}p{105mm}}
Routine && {\tt subroutine dd\_bmeson(i,delta\_mb)} \\
Input && $i=1,2$ - generation index of the lighter valence quark in
the $B^0$ meson, i.e.  $i=2$ chooses $B_s^0$ and $i=1$ chooses
$B_d^0$.\\
Output && ${\tt delta\_mb}=\Delta m_{B_d}$ for $i=1$\\
 && ${\tt delta\_mb}=\Delta m_{B_s}$ for $i=2$\\
QCD related factors:\\
\multicolumn{3}{l} 
{\tt
  common/meson\_data/dmk,amk,epsk,fk,dmd,amd,fd,amb(2),dmb(2),gam\_b(2),fb(2)}
\\
\hskip 9mm Measured $\Delta M_{B_d}^{exp}$ && \hskip 9mm ${\tt
  dmb(1)}= 3.01\cdot 10^{-13}$ \\
\hskip 9mm Measured $\Delta M_{B_s}^{exp}$ && \hskip 9mm ${\tt
  dmb(2)}= 1.2\cdot 10^{-11}$ \\
\hskip 9mm Measured width $\Gamma_{B_d}^{exp}$ && \hskip 9mm ${\tt
  gam\_b(1)}= 1.53\cdot 10^{-12}$ \\
\hskip 9mm Measured width $\Gamma_{B_s}^{exp}$ && \hskip 9mm ${\tt
  gam\_b(1)}= 1.466\cdot 10^{-12}$ \\
\multicolumn{3}{l}
{\tt common/bx\_4q/bk(5),bd(5),bb(2,5),amu\_k,amu\_d,amu\_b}\\
\hskip 9mm $B_1^{\rm VLL} (\mu_B)$ && \hskip 9mm ${\tt bb(1,1)=bb(2,1)}=0.87$\\
\hskip 9mm $B_1^{\rm SLL} (\mu_B)$ && \hskip 9mm ${\tt bb(1,2)=bb(2,2)}=0.8$\\
\hskip 9mm $B_2^{\rm SLL} (\mu_B)$ && \hskip 9mm ${\tt bb(1,3)=bb(2,3)}=0.71$\\
\hskip 9mm $B_1^{\rm LR} (\mu_B)$ && \hskip 9mm ${\tt bb(1,4)=bb(2,4)}=1.71$\\
\hskip 9mm $B_2^{\rm LR} (\mu_B)$ && \hskip 9mm ${\tt bb(1,5)=bb(2,5)}=1.16$\\
\hskip 9mm Renormalization scale $\mu_B$ && \hskip 9mm ${\tt
  amu\_b}=4.6$\\
\multicolumn{3}{l}
{\tt
  common/sm\_4q/eta\_cc,eta\_ct,eta\_tt,eta\_b,bk\_sm,bd\_sm,bb\_sm(2)}\\
\hskip 9mm $B_{SM B_d}^{\rm VLL}$ && \hskip 9mm ${\tt bb\_sm(1)}=1.22$\\
\hskip 9mm $B_{SM B_s}^{\rm VLL}$ && \hskip 9mm ${\tt bb\_sm(2)}=1.22$\\
\hskip 9mm $\eta_b$ && \hskip 9mm ${\tt eta\_b}=0.55$\\
Details of calculations: && Ref.~\cite{BCRS}\\
\end{tabular}

\subsection{$B^0\ra X_s \gamma$ decay rate}
\label{sec:bsgamma}

Both the SUSY contributions and the QCD corrections to the calculation
of the $B^0\ra X_s \gamma$ decay rate are quite complex.  Their
implementation in \code{} is based on the SUSY loop calculations
performed by the authors (not published in a general form) and on the
QCD evolution published in~\cite{CMM98}.  There are no user-accessible
QCD factors apart from the arguments of the {\tt bxg\_nl}
routine.\\[2mm]
\begin{tabular}{lp{5mm}p{95mm}}
Routine && {\tt double precision function bxg\_nl(del,amiu\_b)}\\
Input && {\tt del} - relative photon energy infrared cutoff scale,
$E_\gamma\ge (1- {\tt del}) E_\gamma^{max}$, $0 < {\tt del} < 1$\\
 && {\tt amiu\_b} - renormalization scale \\
Output && $Br(B\ra X_s \gamma)$.\\
Details of calculations: && General SUSY diagrams unpublished, QCD
corrections based on~\cite{CMM98}\\
\end{tabular}

\section{Summary and Outlook}
\label{sec:summary}

We have presented \code, a tool for calculating the set of important
FCNC and CPV observables in the general $R$-parity conserving MSSM.
All implemented physical quantities (listed in Table~\ref{tab:proc})
can be calculated simultaneously for a given set of MSSM parameters.
The calculations of the SUSY particle spectrum and flavor mixing
matrices are performed exactly, so the code can be used for a
completely general pattern of soft SUSY flavor violating terms and
complex phases, without restrictions on the size of sfermion mass
insertions.

Besides complete routines for calculating the physical observables,
\code{} also provides an extensive library of parton-level Green's
functions and Wilson coefficients of many effective quark and lepton
operators (see Table~\ref{tab:green}).  This set actually contains
many more functions than are necessary to compute the quantities
listed in Table~\ref{tab:proc}.  These intermediate building blocks
can be used by \code{} users to construct amplitudes for processes
beyond those already fully implemented by dressing appropriate
combinations of available form factors in QCD corrections and hadronic
matrix elements, without repeating tedious one-loop SUSY calculations
from scratch.  For instance, the form factors implemented in \code{}
for the analysis of $B\ra X_s\gamma$ and $B_{d(s)}\to l^+ l^-$
decays~\cite{MIPORO,DRT} are sufficient to also calculate the $B\to K
l^+ l^-$ decay rate.

\code{} internally uses the conventions of Ref.~\cite{PRD41}, however
in order to facilitate comparison with other programs that analyze
various sectors of MSSM, we have implemented an option to input
parameters in the SLHA2 format~\cite{SLHA2}.

\code{} has been written in FORTRAN 77 and runs fairly quickly; it is
capable of producing a reasonably wide-range scan over the MSSM
parameters within hours or days on a typical personal computer.

The \code{} library is an open project.  We want to gradually add more
features in its future versions.  In particular, we plan to:
\begin{itemize}
\item add more observables in the $B$-meson system, like the CP
  asymmetries in $\bar B B$ meson mixing and in $B\to X_s \gamma$
  decay, as well as observables associated with $B\to K l^+ l^-$
  decay.
\item add observables for lepton flavor-violating processes like
  $\ell^{J}\to \ell^{I}\gamma$, $\ell^{J}\to
  \ell^{K}\ell^{L}\ell^{M}$, and for the lepton anomalous magnetic
  moments, $(g-2)_{I}$
\item include quantities related to FCNCs in the top sector, like
  $t\to c X$ with $X=\gamma, Z, g, H$, in order to probe the flavor
  violation in up-squark mass matrices that are (almost) unconstrained
  to this moment.
\item implement full resummation of leading large $\tan\beta$ effects
  beyond the MFV scenario.
\end{itemize}

With the increasing accuracy of experimental data on flavor and CP
violation in rare processes, it may eventually become possible to not
only constrain the MSSM parameters, but also, if significant
deviations from the SM predictions are found, to recover their actual
values.  For that multi-process analysis, such as the one performed by
\code, will be necessary.  Therefore, we hope that \code{} becomes an
important tool that is useful not only to theorists working on MSSM
but also to experimentalists fitting the MSSM onto forthcoming data
from the Tevatron, LHC, and $B$-factories.

\subsection*{Acknowledgments}

\noindent Authors would like to thank A.~Buras, T.~Ewerth, M.~Misiak,
C.~Savoy, {\L}.~Slawianowska and S.~Pokorski for collaboration in
performing theoretical calculations used in \code{} and for helping to
check and debug some of its sections.  We would also like to thank
W. Altmannshofer, D.  Guadagnoli and M. Wick for careful checking the
$Br(B\ra X_s \gamma)$ code and reporting some inconsistencies.

This work is supported by the RTN European Programme,
MRTN-CT-2006-035505 (HEPTOOLS, Tools and Precision Calculations for
Physics Discoveries at Colliders).  JR was also supported in part the
Polish Ministry of Science and Higher Education Grant N~N202~230337.
J.R.  and A.D.  acknowledge partial support by the EU FP6 Marie Curie
Research and Training Network ``UniverseNet" (MRTN-CT-2006-035863).
P.T.  was supported by a Marshall Scholarship and a National Science
Foundation Graduate Research Fellowship.


\def\theequation{\Alph{section}.\arabic{equation}}
\begin{appendix}

\setcounter{equation}{0}

\parindent 0cm

\section{Installation of the program}
\label{app:inst}

The installation and execution of \code{} is very simple.  On Unix or
Linux systems, just follow these steps :

\begin{enumerate}

\item Download the code from \webpage{} and unpack it.

\item Change directory into {\tt susy\_{flavor}}.

\item Edit {\tt Makefile} and change {\tt F77 = gfortran} and {\tt
  FOPT = -O -fno-automatic -Wall} into your compiler name and options,
  respectively.

\item Exit {\tt Makefile} and type {\tt make} (or {\tt gmake}).

\item If everything go through the code will ask you whether to read
  the input file {\tt susy\_flavor.in} or to use the parameters
  defined inside the driver file.

\item To run the code from now on just type {\tt ./sflav}.

\end{enumerate}

The authors tested \code{} on Linux machines.  With few
straightforward modifications the procedure describe above can be
adapted to install program on other systems.

The output of the program is displayed on the screen.  In addition a
file named {\tt mssm\_data.txt} is created.  It contains information
about the MSSM Lagrangian parameters and the tree-level mass spectrum
corresponding to the input parameter set.  A sample set of input
parameters and corresponding \code{} output are listed in the
following appendices.

\section{Example of the \code{} initialization sequence}
\label{app:code}

Below we present the contents of {\tt susy\_flavor.f}, the master
driver file for the \code{} library.  The driver program illustrates
the correct initialization sequence for all relevant MSSM parameters
(see Section~\ref{sec:init}) and shows how to perform calls to the
routines calculating physical observables (Section~\ref{sec:proc}).

The driver file asks if the input parameters should be given directly
inside the program or read from the default input file named {\tt
  susy\_flavor.in} (in this case skipping the values given in the
program).  Defining the input parameters in the separate file is
probably more straightforward, but the ability to initialize
parameters from within the program could be more useful for performing
multi-dimensional scans over the MSSM parameter space.

\begin{small}

{\tt
\begin{tabbing}
~~~~~~~\=program susy\_flavor ~~~~~~~~~~ \= \+ \\
      implicit double precision (a-h,o-z)\\
      dimension sll(3),slr(3),sql(3),squ(3),sqd(3)\\
      double complex asl(3),asu(3),asd(3)\\
      double complex slmi\_l(3),slmi\_r(3),slmi\_lr(3,3)\\
      double complex sqmi\_l(3),sdmi\_r(3),sumi\_r(3)\\
      double complex sdmi\_lr(3,3),sumi\_lr(3,3)\\
      double complex amg,amgg,amue\\
      common/sf\_cont/eps,indx(3,3),iconv\\
\- \\
{\it c} \> \+ {\it decide if input parameters are read from file
  susy\_flavor.in or defined inside the program}\\
      write(*,'(a,\$)')'Read input from file susy\_flavor.in
      (no=1,yes=2)? '\\
      read(*,*) input\_type\\
      if (input\_type.eq.2) then \\
         ~~call sflav\_input \> {\it ! Parameters read from file
        susy\_flavor.in }\\
         ~~goto 100\\
      end if\\
\- \\
{\it c} \> \+ {\it Parameters defined inside the code.  Start from
  input convention choice} \- \\
{\it c} \> \+ {\it iconv = 1} \> {\it !  SLHA2 input conventions}\\
      iconv = 2 \> {\it !  hep-ph/9511250 input conventions}\-\\
\\
{\it c} \> \+ {\it SM basic input initialization}\\
      zm0 = 91.1876d0 \> {\it !  M\_Z}\\
      wm0 = 80.398d0 \> {\it !  M\_W}\\
      alpha\_z = 1/127.934d0 \> {\it !  alpha\_em(M\_Z)}\\
      call vpar\_update(zm0,wm0,alpha\_z) \-\\
\\
{\it c} \> \+ {\it QCD parameters}\\
      alpha\_s = 0.1172d0 \> {\it !  alpha\_s(M\_Z)}\\
      call lam\_fit(alpha\_s) \> {\it ! fits Lambda\_QCD at 3 loop
        level}\\
      call lam\_fit\_nlo(alpha\_s) \> {\it !  fits Lambda\_QCD at NLO
        level} \-\\
\\
{\it c} \> \+ {\it CKM matrix initialization}\\
      alam = 0.2258d0 \> {\it !  lambda}\\
      apar = 0.808d0 \> {\it !  A}\\
      rhobar = 0.177d0 \> {\it !  rho bar}\\
      etabar = 0.360d0 \> {\it !  eta bar}\\
      call ckm\_wolf(alam,apar,rhobar,etabar) \- \\
\\
{\it c} \> \+ {\it Fermion mass initialization, input: MSbar running
  quark masses}\\
      top\_scale = 163.2d0\\
      top = 163.2d0 \> {\it !  m\_t(top\_scale)}\\
      bot\_scale = 4.17d0\\
      bot = 4.17d0 \> {\it !  m\_b(bot\_scale)}\\
      call init\_fermion\_sector(top,top\_scale,bot,bot\_scale) \-\\
\\
{\it c} \> \+ {\it Higgs sector parameters}\\
      pm = 200 \> {\it !  M\_A}\\
      tanbe = 10 \> {\it !  tan(beta)}\\
      amue = (200.d0,100.d0) \> {\it !  mu parameter}\\
      call init\_higgs\_sector(pm,tanbe,amue,ierr)\\
      if (ierr.ne.0) stop 'negative tree level Higgs mass$^2$?' \- \\
\\
{\it c} \> \+ {\it Gaugino sector parameters: if M1=0 set here then
  program uses M1 = $5s_W^2/3c_W^2$ M2}\\
      amgg = (0.d0,0.d0) \> {\it !  M1 (bino mass, complex)}\\
      amg = (200.d0,0.d0) \> {\it !  M2 (wino mass, complex)}\\
      amglu = 3*abs(amg) \> {\it !  M3 (gluino mass)}\\
      call init\_ino\_sector(amgg,amg,amglu,amue,tanbe,ierr)\\
      if (ierr.ne.0) write(*,*) '-ino mass below M\_Z/2?'\\
\- \\
{\it c} \> \+ {\it Slepton diagonal soft breaking parameters}\\
      sll(1) = 300.d0 \> {\it ! left selectron mass scale}\\
      sll(2) = 300.d0 \> {\it ! left smuon mass scale}\\
      sll(3) = 300.d0 \> {\it ! left stau mass scale}\\
      slr(1) = 300.d0 \> {\it ! right selectron mass scale}\\
      slr(2) = 300.d0 \> {\it ! right smuon mass scale}\\
      slr(3) = 300.d0 \> {\it ! right stau mass scale} \- \\
{\it c} \> \+ {\it Dimensionless (normalized to masses) slepton
  diagonal LR mixing } \\
      asl(1) = (1.d0,0.d0) \> {\it ! 1st generation} \\
      asl(2) = (1.d0,0.d0) \> {\it ! 2nd generation} \\
      asl(3) = (1.d0,0.d0) \> {\it ! 3rd generation} \- \\
{\it c} \> \+ {\it Slepton LL and RR mass insertions (hermitian
  matrices, only upper part given)}\- \\
{\it c} \> \+ {\it slmi\_x(1),slmi\_x(2), slmi\_x(3) are 12,23,31
  entry, respectively}\\
      do i=1,3\\
         ~~slmi\_l(i) = (0.d0,0.d0) \> {\it !  slepton LL mass
           insertion}\\
         ~~slmi\_r(i) = (0.d0,0.d0) \> {\it !  slepton RR mass
           insertion}\\
      end do \\
      slmi\_l(2) = (2.d-2,1.d-2) \> {\it !  example, non-vanishing LL
        23 entry}\- \\
{\it c} \> \+ {\it Slepton LR mass insertions, non-hermitian in
  general}\\
      do i=1,3\\
         ~~do j=1,3\\
            ~~~~slmi\_lr(i,j) = (0.d0,0.d0) \> {\it !  slepton LR ij mass
              insertion}\\
         ~~end do\\
      end do \- \\
{\it c} \> \+ {\it Calculate slepton physical masses and mixing
  angles}\\
      call init\_slepton\_sector(sll,slr,asl,ierr,slmi\_l,slmi\_r,slmi\_lr)\\
      if (ierr.ne.0) stop 'negative tree level slepton mass$^2$?'\\
\- \\
{\it c} \> \+ {\it Squark diagonal soft breaking parameters}\\
      sql(1) = 500.d0 \> {\it ! left squark mass, 1st
        generation}\\
      sql(2) = 500.d0 \> {\it ! left squark mass, 2nd
        generation}\\
      sql(3) = 400.d0 \> {\it ! left squark mass, 3rd
        generation}\\
      sqd(1) = 550.d0 \> {\it ! right down squark mass}\\
      sqd(2) = 550.d0 \> {\it ! right strange squark mass}\\
      sqd(3) = 300.d0 \> {\it ! right sbottom mass}\\
      squ(1) = 450.d0 \> {\it ! right up squark mass}\\
      squ(2) = 450.d0 \> {\it ! right charm squark mass}\\
      squ(3) = 200.d0 \> {\it ! right stop mass}\-\\
{\it c} \> \+ {\it Dimensionless (normalized to masses) squark
  diagonal LR mixing} \\
      asd(1) = (1.d0,0.d0) \> {\it ! down squark LR mixing, 1st
        generation}\\
      asd(2) = (1.d0,0.d0) \> {\it ! down squark LR mixing, 2nd
        generation}\\
      asd(3) = (1.d0,0.d0) \> {\it ! down squark LR mixing, 3rd
        generation}\\
      asu(1) = (1.d0,0.d0) \> {\it ! up squark LR mixing, 1st
        generation}\\
      asu(2) = (1.d0,0.d0) \> {\it ! up squark LR mixing, 2nd
        generation}\\
      asu(3) = (1.d0,0.d0) \> {\it ! up squark LR mixing, 3rd
        generation}\-\\
{\it c} \> \+ {\it Squark LL and RR mass insertions (hermitian
  matrices, only upper part given)}\- \\
{\it c} \> \+ {\it sqmi\_l(1),sqmi\_l(2), sqmi\_l(3) are 12,23,31
  entry, respectively, etc.}\\
      do i=1,3\\
         ~~sqmi\_l(i) = (0.d0,0.d0) \> {\it !  squark LL mass
        insertion}\\
         ~~sumi\_r(i) = (0.d0,0.d0) \> {\it !  up-squark RR mass
        insertion}\\
         ~~sdmi\_r(i) = (0.d0,0.d0) \> {\it !  down-squark RR mass
        insertion}\\
      end do \\
      sqmi\_l(2) = (2.d-2,-1.d-2) \> {\it !  example, non-vanishing LL
        23 entry}\- \\
{\it c} \> \+ {\it Squark LR mass insertions, non-hermitian in
  general}\\
      do i=1,3\\
         ~~do j=1,3\\
            ~~~~sumi\_lr(i,j) = (0.d0,0.d0) \> {\it !  up-squark LR ij
              mass insertion}\\
            ~~~~sdmi\_lr(i,j) = (0.d0,0.d0) \> {\it !  down-squark LR ij
              mass insertion}\\
         ~~end do\\
      end do \- \\
{\it c} \> \+ {\it Calculate squark physical masses and mixing
  angles}\\
      call
      init\_squark\_sector(sql,squ,sqd,asu,asd,ierr,sqmi\_l,sumi\_r,
      \- \\
~~~~~~\$~~~~ sdmi\_r,sumi\_lr,sdmi\_lr) \+ \\
      if (ierr.ne.0) stop 'negative tree level squark mass$^2$?'\\
\- \\
{\it c} \> \+ {\it reset status of physical Higgs mass after parameter
  changes}\\
      call reset\_phys\_data\- \\
{\it c} \> \+ {\it Neutral CP-even Higgs masses in the 1-loop
  Effective Potential Approximation.}\- \\
{\it c} \> \+ {\it Only real mu, A\_t, A\_b allowed - replace
  x$\ra$abs(x)}\\
      call
      fcorr\_EPA(tanbe,pm,top,abs(amue),sql(3),sqd(3),squ(3),
      \- \\
~~~~~~\$~~~~ abs(asd(3)),abs(asu(3)),ierr) \+ \\
      if (ierr.ne.0) stop 'negative 1-loop EPA CP-even Higgs
      mass$^2$?' \- \\
\\
100 \> continue \> {\it !!! End of input section !!!} \\
{\it c} \> \+ {\it Control output: Lagrangian parameters and tree
  level masses written on file mssm\_data.txt}  \\
      ifl = 1 \> {\it !  output file number}\\
      open(ifl,file='mssm\_data.txt',status='unknown') \\
      call print\_MSSM\_par(ifl) \> {\it ! Lagrangian parameters } \\
      call print\_MSSM\_masses(ifl) \> {\it ! tree level physical
        masses }\\
      close(ifl)\\
\- \\
{\it c} \> \+ {\it Results for implemented rare decays:}\\
      write(*,99)'Electric dipole moments:'\\
      write(*,99)'Electron EDM = ',edm\_l(1)\\
      write(*,99)'Muon EDM     = ',edm\_l(2)\\
      write(*,99)'Tau EDM      = ',edm\_l(3)\\
      write(*,99)'Neutron EDM  = ',edm\_n()\\
\\
      write(*,99)'Neutrino K decays:'\\
      call k\_pivv(br\_k0,br\_kp)\\
      write(*,99)'BR(K$_L^0$ $\ra$ pi$^0$ vv) = ',br\_k0\\
      write(*,99)'BR(K$^+$   $\ra$ pi$^+$ vv) = ',br\_kp\\
\\
      write(*,99)'Leptonic B decays:'\\
      write(*,99)'BR(B\_d $\ra$ mu$^+$ mu$^-$) = ',b\_ll(3,1,2,2)\\
      write(*,99)'BR(B\_s $\ra$ mu$^+$ mu$^-$) = ',b\_ll(3,2,2,2)\\
\\
      write(*,99)'B$\ra$ X\_s photon decay:'\- \\
{\it c} \> \+ {\it Physical quantities for BR(B$\ra$X\_s g) calculation}
\\
      delb = 0.99d0 \> {\it !  Photon energy infrared cutoff}\\
      amiu\_b= 4.8d0 \> {\it !  Renormalization scale miu\_b}\\
      write(*,99)'BR(B $\ra$ X\_S gamma) = ',bxg\_nl(delb,amiu\_b)\\
\\
      write(*,99)'KK mixing:'\\
      call dd\_kaon(eps\_k,delta\_mk)\\
      write(*,99)'eps\_K = ',eps\_k\\
      write(*,99)'Delta m\_K = ',delta\_mk\\
\\
      write(*,99)'DD mixing:'\\
      call uu\_dmeson(delta\_md)\\
      write(*,99)'Delta m\_D = ',delta\_md\\
\\
      write(*,99)'BB mixing:'\\
      call dd\_bmeson(1,delta\_mbd)\\
      write(*,99)'Delta m\_B\_d = ',delta\_mbd\\
      call dd\_bmeson(2,delta\_mbs)\\
      write(*,99)'Delta m\_B\_s = ',delta\_mbs\\
\\
 99   format(a,1pe11.4)\\
      end
\end{tabbing}
}
\end{small}

\section{Example of  \code{} input file}
\label{app:infile}

By default, the driver file {\tt susy\_flavor.f} reads input
parameters from the file {\tt susy\_flavor.in}.  Below we provide an
example input file defining a set of parameters equivalent to those in
the driver file presented in Appendix~\ref{app:code}.
\begin{small}
{\tt

\begin{tabbing}

~~~~\=~~~~\=~~~~~~~~~~~~~~~~~~~~~~~~~~~~\=~~~~~\=\\

\# Example input of SUSY\_FLAVOR in Les Houches Accord-like format\\
\#\\
\# CAUTION: users can modify numerical data in this file but they\\
\# should not remove existing data lines within blocks SMINPUTS,\\
\# VCKMIN, EXTPAR, MSL2IN, MSE2IN, MSQ2IN, MSU2IN, MSD2IN, TEIN, TUIN,\\
\# TDIN, IMMSL2IN, IMMSE2IN, IMMSQ2IN, IMMSU2IN, IMMSD2IN, IMTEIN,\\
\# IMTUIN, IMTDIN.  New data lines in each block can be added but only\\
\# after the already defined ones.  Also, comment-only lines starting\\
\# from \# as a first character can be added only just after or before\\
\# Block XXX statements, i.e. not between data lines. Otherwise\\
\# SUSY\_FLAVOR input routine sflav\_input will denounce input file as\\
\# corrupted or read incorrect values.\\
\#\\
\# Full new data blocks can be added, sflav\_input will ignore them.\\
\# \\
Block MODSEL  \> \> \>   \# Select model\\
\>    1 \>   0	\>     \# General MSSM\\
\>    3	\>   0	\>     \# MSSM particle content\\
\>    4	\>   0	\>     \# R-parity conserving MSSM\\
\>    5	\>   2	\>     \# CP violated\\
\>    6	\>   3	\>     \# Lepton and quark flavor violated\\
Block SOFTINP	\> \> \>     \# Choose convention for the soft terms\\
\# convention = 1:\\
\# \>  sfermion input parameters in SLHA2 conventions \\
\# convention = 2:\\
\# \>  sfermion input parameters in conventions of hep-ph/9511250\\
\# input\_type = 1:\\
\# \> sfermion off-diagonal terms given as dimensionless mass insertions\\
\# \>   LR diagonal terms given as dimensionless parameters\\
\# input\_type = 2:\\
\# \>  sfermion soft terms given as absolute values\\
\# See comment in Blocks MSXIN2, TXIN below\\
\>    1 \> 2   \>  \# sfermion convention, SLHA2 or hep-ph/9511250\\
\>    2 \> 1 \>    \# input\_type (dimension of soft mass entries) \\
Block SMINPUTS  \> \> \>  \# Standard Model inputs\\
\>    1 \> 1.279340000e+02   \>  \# alpha$^{-1}$ SM MSbar(MZ)\\
\>    3 \>  1.172000000e-01  \>   \# alpha\_s(MZ) SM MSbar\\
\>    4 \>  9.118760000e+01  \>   \# MZ(pole)\\
\>    5 \>  4.170000000e+00   \>  \# mb(mb) SM MSbar\\
\>    6 \>  1.632000000e+02  \>   \# mtop(mt) SM MSbar\\
\>    7 \>  1.777000000e+00  \>   \# mtau(pole)\\
\>    11 \> 5.110000000e-04  \>   \# me(pole)\\
\>    13 \> 1.056590000e-01  \>    \# mmu(pole)\\
\>    21 \> 7.000000000e-03  \>   \# md(2 GeV) MSbar\\
\>    22 \> 4.000000000e-03  \>   \# mu(2 GeV) MSbar\\
\>    23 \> 1.100000000e-01  \>   \# ms(2 GeV) MSbar\\
\>    24 \> 1.279000000e+00   \>  \# mc(mc) MSbar\\
\>    30 \> 8.039800000e+01  \>  \# MW (pole), not standard SLHA2 entry!!!\\
Block VCKMIN  \> \> \>   \# CKM matrix\\
\>    1 \>  2.258000000e-01  \>   \# lambda\\
\>    2 \>  8.080000000e-01  \>   \# A\\
\>    3 \>  1.770000000e-01  \>   \# rho bar\\
\>    4 \>  3.600000000e-01  \>   \# eta bar\\
Block EXTPAR   \> \> \>          \# non-minimal input parameters, real part\\
\>    1 \>  0.000000000e+02 \>     \# Re(m1), U(1) gaugino mass\\
\>    2 \>  2.000000000e+02 \>     \# Re(m2), SU(2) gaugino mass\\
\>    3 \>  6.000000000e+02 \>     \# m3, SU(3) gaugino mass\\
\>    23\>	2.000000000e+02	 \>    \# Re(mu)\\ 
\>    25\>	1.000000000e+01	 \>    \# tan(beta)\\
\>    26\>	2.000000000e+02	 \>    \# MA\\
Block IMEXTPAR     \> \> \>   \# non-minimal input parameters, imaginary part\\
\>    1 \>  0.000000000e+00 \>     \# Im(m1), U(1) gaugino mass\\
\>    2 \>  0.000000000e+00   \>   \# Im(m2), SU(2) gaugino mass\\
\>    23\>	1.000000000e+02  \>    \# Im(mu)\\
\# if abs(m1) = 0 SUSY\_FLAVOR uses m1=5/3 s\_W$^2$/c\_W$^2$ m2\\
\#   \\
\# Soft sfermion mass matrices\\
\#\\
\# Off-diagonal entries may be given as absolute entries or as\\
\# dimensionless mass insertions - then real off-diagonal entries of\\
\# SLHA2 blocks are calculated by SUSY\_FLAVOUR as\\
\# M$^2$(I,J) = (mass insertion)(I,J) sqrt(M$^2$(I,I) M$^2$(J,J))\\
\# (see comments at the top of subroutine sflav\_input) \\
\#\\
\# Below we give an example of dimensionless off-diagonal entries\\
\#\\
Block MSL2IN   \> \> \>                  \# left soft slepton mass matrix, real part\\
\> 1  1 \> 9.000000000e+04 \>    \# Left slepton diagonal mass$^2$, 1st generation\\
\> 2  2 \> 9.000000000e+04  \>     \# Left slepton diagonal mass$^2$, 2nd generation\\
\> 3  3 \> 9.000000000e+04  \>    \# Left slepton diagonal mass$^2$, 3rd generation\\
\> 1  2 \> 0.000000000e+00  \>     \# Dimensionless left slepton mass insertion 12\\
\> 2  3 \> 2.000000000e-02  \>     \# Dimensionless left slepton mass insertion 23\\
\> 1  3 \> 0.000000000e+00  \>     \# Dimensionless left slepton mass insertion 13\\
Block IMMSL2IN    \> \> \>               \# left soft slepton mass matrix, imaginary part\\
\> 1  2\>  0.000000000e+00    \>   \# Dimensionless left slepton mass insertion 12\\
\> 2  3\>  1.000000000e-02   \>    \# Dimensionless left slepton mass insertion 23\\
\> 1  3 \> 0.000000000e+00   \>    \# Dimensionless left slepton mass insertion 13\\
Block MSE2IN      \> \> \>               \# right soft slepton mass matrix, real part\\
\> 1  1\>  9.000000000e+04   \>   \# Right selectron diagonal mass$^2$\\
\> 2  2 \> 9.000000000e+04   \>   \# Right smuon diagonal mass$^2$\\
\> 3  3 \> 9.000000000e+04   \>   \# Right stau diagonal mass$^2$\\
\> 1  2\>  0.000000000e+00   \>    \# Dimensionless right slepton mass insertion 12\\
\> 2  3\>  0.000000000e+00   \>    \# Dimensionless right slepton mass insertion 23\\
\> 1  3\>  0.000000000e+00   \>    \# Dimensionless right slepton mass insertion 13\\
Block IMMSE2IN     \> \> \>              \# right soft slepton mass matrix, imaginary part\\
\> 1  2\>  0.000000000e+00   \>   \# Dimensionless right slepton mass insertion 12\\
\> 2  3\>  0.000000000e+00   \>    \# Dimensionless right slepton mass insertion 23\\
\> 1  3\>  0.000000000e+00   \>    \# Dimensionless right slepton mass insertion 13\\
Block MSQ2IN        \> \> \>             \# left soft squark mass matrix, real part\\
\> 1  1\>  2.500000000e+05   \>   \# Left squark diagonal mass$^2$, 1st generation\\
\> 2  2\>  2.500000000e+05   \>   \# Left squark diagonal mass$^2$, 2nd generation\\
\> 3  3\>  1.600000000e+05   \>   \# Left squark diagonal mass$^2$, 3rd generation\\
\> 1  2\>  0.000000000e+00   \>    \# Dimensionless left squark mass insertion 12\\
\> 2  3\>  2.000000000e-02   \>    \# Dimensionless left squark mass insertion 23\\
\> 1  3 \> 0.000000000e+00   \>    \# Dimensionless left squark mass insertion 13\\
Block IMMSQ2IN   \> \> \>                \# left soft squark mass matrix, imaginary part\\
\> 1  2\>  0.000000000e+00    \>   \# Dimensionless left squark mass insertion 12\\
\> 2  3 \>-1.000000000e-02    \>   \# Dimensionless left squark mass insertion 23\\
\> 1  3 \> 0.000000000e+00    \>   \# Dimensionless left squark mass insertion 13\\
Block MSU2IN     \> \> \>                \# right soft up-squark mass matrix, real part\\
\> 1  1 \> 2.025000000e+05  \>    \# Right u-squark diagonal mass$^2$\\
\> 2  2 \> 2.025000000e+05  \>    \# Right c-squark diagonal mass$^2$\\
\> 3  3 \> 4.000000000e+04  \>    \# Right stop diagonal mass$^2$\\
\> 1  2 \> 0.000000000e+00  \>     \# Dimensionless right up-squark mass insertion 12\\
\> 2  3 \> 0.000000000e+00  \>     \# Dimensionless right up-squark mass insertion 23\\
\> 1  3 \> 0.000000000e+00   \>    \# Dimensionless right up-squark mass insertion 13\\
Block IMMSU2IN   \> \> \>                \# right soft up-squark mass matrix, imaginary part\\
\> 1  2 \> 0.000000000e+00   \>    \# Dimensionless right up-squark mass insertion 12\\
\> 2  3\>  0.000000000e+00   \>    \# Dimensionless right up-squark mass insertion 23\\
\> 1  3 \> 0.000000000e+00   \>    \# Dimensionless right up-squark mass insertion 13\\
Block MSD2IN                 \# right soft down-squark mass matrix, real part\\
\> 1  1 \> 3.025000000e+05   \>   \# Right d-squark diagonal mass$^2$\\
\> 2  2\>  3.025000000e+05   \>   \# Right s-squark diagonal mass$^2$\\
\> 3  3 \> 9.000000000e+04   \>   \# Right sbottom diagonal mass$^2$\\
\> 1  2 \> 0.000000000e+00   \>    \# Dimensionless right down-squark mass insertion 12\\
\> 2  3 \> 0.000000000e+00   \>    \# Dimensionless right down-squark mass insertion 23\\
\> 1  3 \> 0.000000000e+00   \>    \# Dimensionless right down-squark mass insertion 13\\
Block IMMSD2IN     \> \> \>              \# right soft down-squark mass matrix, imaginary part\\
\> 1  2 \> 0.000000000e+00   \>    \# Dimensionless right down-squark mass insertion 12\\
\> 2  3 \> 0.000000000e+00   \>    \# Dimensionless right down-squark mass insertion 23\\
\> 1  3 \> 0.000000000e+00   \>    \# Dimensionless right down-squark mass insertion 13\\
\#\\
\# Soft sfermion trilinear mixing matrices\\
\#\\
\# LR mixing parameters can be given as absolute entries or as\\
\# dimensionless diagonal A-terms and dimensionless ff-diagonal mass\\
\# insertions - see comments at the top of subroutine sflav\_input\\
\#\\
\# Below we give an example of dimensionless A terms.\\
\#\\
\# Diagonal entries below are dimensionless "A parameters"\\
\# Diagonal entries of SLHA2 LR blocks are calculated by SUSY\_FLAVOUR as\\
\# TL(I,I) = AL(I,I) Yukawa\_L(I) sqrt(ML$^2$(I,I)*ME$^2$(I,I))\\
\# TU(I,I) = AU(I,I) Yukawa\_U(I) sqrt(MQ$^2$(I,I)*MU$^2$(I,I))\\
\# TD(I,I) = AD(I,I) Yukawa\_D(I) sqrt(MQ$^2$(I,I)*MD$^2$(I,I))\\
\#\\
\# Off-diagonal entries are dimensionless "mass insertions"\\
\# Off-diagonal entries of SLHA2 LR blocks are calculated by SUSY\_FLAVOUR as\\
\#\\
\# TL(I,J) = AL(I,J) sqrt(2 ML$^2$(I,I)*ME$^2$(J,J))/v1\\
\# TU(I,J) = AU(I,J) sqrt(2 MQ$^2$(I,I)*MU$^2$(J,J))/v2\\
\# TD(I,J) = AD(I,J) sqrt(2 MQ$^2$(I,I)*MD$^2$(J,J))/v1\\
\#\\
Block TEIN    \> \> \>             \# slepton trilinear mixing, dimensionless, real part\\
\> 1  1 \> 1.000000000e+00\>	     \# Diagonal AL term, 1st generation\\
\> 2  2 \> 1.000000000e+00 \>      \# Diagonal AL term, 2nd generation\\
\> 3  3 \> 1.000000000e+00 \>      \# Diagonal AL term, 3rd generation\\
\> 1  2 \> 0.000000000e+00 \>      \# Slepton LR mass insertion 12\\
\> 2  1 \> 0.000000000e+00 \>      \# Slepton LR mass insertion 21\\
\> 2  3 \> 0.000000000e+00 \>      \# Slepton LR mass insertion 23\\
\> 3  2 \> 0.000000000e+00 \>      \# Slepton LR mass insertion 32\\
\> 1  3 \> 0.000000000e+00 \>      \# Slepton LR mass insertion 13\\
\> 3  1 \> 0.000000000e+00  \>     \# Slepton LR mass insertion 31\\
Block IMTEIN    \>\>\>             \# slepton trilinear mixing, dimensionless, imag. part\\
\> 1  1\>  0.000000000e+00\>	     \# Diagonal AL term, 1st generation\\
\> 2  2\>  0.000000000e+00  \>     \# Diagonal AL term, 2nd generation\\
\> 3  3 \> 0.000000000e+00  \>     \# Diagonal AL term, 3rd generation\\
\> 1  2 \> 0.000000000e+00  \>     \# Slepton LR mass insertion 12\\
\> 2  1 \> 0.000000000e+00  \>     \# Slepton LR mass insertion 21\\
\> 2  3 \> 0.000000000e+00  \>     \# Slepton LR mass insertion 23\\
\> 3  2 \> 0.000000000e+00  \>     \# Slepton LR mass insertion 32\\
\> 1  3 \> 0.000000000e+00  \>     \# Slepton LR mass insertion 13\\
\> 3  1 \> 0.000000000e+00  \>     \# Slepton LR mass insertion 31\\
Block TUIN    \>\>\>               \# up-squark trilinear mixing, dimensionless, real part\\
\> 1  1 \> 1.000000000e+00\>	     \# Diagonal AU term, 1st generation\\
\> 2  2 \> 1.000000000e+00\>	     \# Diagonal AU term, 2nd generation\\
\> 3  3 \> 1.000000000e+00\>	     \# Diagonal AU term, 3rd generation\\
\> 1  2 \> 0.000000000e+00 \>      \# Up-squark LR mass insertion 12\\
\> 2  1\>  0.000000000e+00 \>      \# Up-squark LR mass insertion 21\\
\> 2  3 \> 0.000000000e+00 \>      \# Up-squark LR mass insertion 23\\
\> 3  2 \> 0.000000000e+00  \>     \# Up-squark LR mass insertion 32\\
\> 1  3\>  0.000000000e+00  \>     \# Up-squark LR mass insertion 13\\
\> 3  1 \> 0.000000000e+00  \>     \# Up-squark LR mass insertion 31\\
Block IMTUIN      \>\>\>           \# up-squark trilinear mixing, dimensionless, imag. part\\
\> 1  1 \> 0.000000000e+00\>	     \# Diagonal AU term, 1st generation\\
\> 2  2 \> 0.000000000e+00\>	     \# Diagonal AU term, 2nd generation\\
\> 3  3 \> 0.000000000e+00\>	     \# Diagonal AU term, 3rd generation\\
\> 1  2 \> 0.000000000e+00   \>    \# Up-squark LR mass insertion 12\\
\> 2  1 \> 0.000000000e+00   \>    \# Up-squark LR mass insertion 21\\
\> 2  3 \> 0.000000000e+00    \>   \# Up-squark LR mass insertion 23\\
\> 3  2 \> 0.000000000e+00   \>    \# Up-squark LR mass insertion 32\\
\> 1  3 \> 0.000000000e+00   \>    \# Up-squark LR mass insertion 13\\
\> 3  1 \> 0.000000000e+00    \>   \# Up-squark LR mass insertion 31\\
Block TDIN   \>\>\>                \# down-squark trilinear mixing, dimensionless, real part\\
\> 1  1\>  1.000000000e+00\>	     \# Diagonal AD term, 1st generation\\
\> 2  2\>  1.000000000e+00\>	     \# Diagonal AD term, 2nd generation\\
\> 3  3 \> 1.000000000e+00\>	     \# Diagonal AD term, 3rd generation\\
\> 1  2 \> 0.000000000e+00  \>     \# Down-squark LR mass insertion 12\\
\> 2  1 \> 0.000000000e+00  \>     \# Down-squark LR mass insertion 21\\
\> 2  3 \> 0.000000000e+00 \>      \# Down-squark LR mass insertion 23\\
\> 3  2 \> 0.000000000e+00 \>      \# Down-squark LR mass insertion 32\\
\> 1  3 \> 0.000000000e+00 \>      \# Down-squark LR mass insertion 13\\
\> 3  1\>  0.000000000e+00 \>      \# Down-squark LR mass insertion 31\\
Block IMTDIN  \>\>\>             \# down-squark trilinear mixing, dimensionless, imag. part\\
\> 1  1\>  0.000000000e+00\>	     \# Diagonal AD term, 1st generation\\
\> 2  2\>  0.000000000e+00\>	     \# Diagonal AD term, 2nd generation\\
\> 3  3 \> 0.000000000e+00\>	     \# Diagonal AD term, 3rd generation\\
\> 1  2 \> 0.000000000e+00 \>      \# Down-squark LR mass insertion 12\\
\> 2  1\>  0.000000000e+00 \>      \# Down-squark LR mass insertion 21\\
\> 2  3 \> 0.000000000e+00 \>      \# Down-squark LR mass insertion 23\\
\> 3  2\>  0.000000000e+00 \>      \# Down-squark LR mass insertion 32\\
\> 1  3\>  0.000000000e+00 \>      \# Down-squark LR mass insertion 13\\
\> 3  1 \> 0.000000000e+00 \>      \# Down-squark LR mass insertion 31\\
\end{tabbing}
}
\end{small}

\section{Example of  \code{} output}
\label{app:output}

The parameters defined inside the driver program in
Appendix~\ref{app:code} and in the input file listed in
Appendix~\ref{app:infile} should produce identical output.  We enclose
it here so that \code{} users can check that the program gives the
same result on their own computers and FORTRAN compilers.

The driver file {\tt susy\_flavor.f} writes the MSSM Lagrangian
parameters and tree-level particle masses to the file {\tt
  mssm\_data.txt}.  For the parameters defined in
Appendices~\ref{app:code} and~\ref{app:infile} one has:
\begin{small} 

{\tt 

\begin{tabbing}
~~~~~~~~~~~~~~~~~~~~~~~~~~~~~~~~~~~~~\=~~~~~~~~~~~~~~~~\= \\[-2mm]
******* MSSM Lagrangian parameters *******\\
QED coupling 1/alpha\_em(M\_Z) \> =   1.2793E+02\\
Weinberg angle s\_W$^2$ \> =  2.2265E-01\\
Z boson mass \> =  9.1188E+01\\
W boson mass \> =  8.0398E+01\\
QCD coupling alpha\_s(M\_Z) \> =  1.1720E-01\\
\\
Higgs mixing parameter mu (complex)  \> =  2.0000E+02 \>  1.0000E+02\\
Higgs soft mixing parameter m$_{12}^2$ \>  = -3.9604E+03\\
Higgs soft masses m$_{H_1}^2$,m$_{H_2}^2$ \> = -6.3208E+03 \> -5.3679E+04\\
\\
U(1)  gaugino mass (complex) \> = 9.5472E+01 \>  0.0000E+00\\
SU(2) gaugino mass (complex) \> = 2.0000E+02 \>  0.0000E+00\\
SU(3) gaugino mass (real)    \> = 6.0000E+02\\
~~~~~~~~~~~~~~~~\=~~~~~~~~~~~~~~~~\=  \\
Left slepton mass matrix, real part:\\
 9.00009E+04 \>  0.00000E+00  \> 0.00000E+00\\
 0.00000E+00 \>  9.00018E+04 \>  1.80004E+03\\
 0.00000E+00 \>  1.80004E+03  \> 9.00027E+04\\
Left slepton mass matrix, imaginary part:\\
 0.00000E+00 \>  0.00000E+00 \>  0.00000E+00\\
-0.00000E+00 \>  0.00000E+00  \> 9.00022E+02\\
-0.00000E+00 \> -9.00022E+02  \> 0.00000E+00\\
\\
Right slepton mass matrix, real part:\\
 8.99991E+04 \>  0.00000E+00  \> 0.00000E+00\\
 0.00000E+00 \>  8.99982E+04 \>  0.00000E+00\\
 0.00000E+00 \>  0.00000E+00  \> 8.99973E+04\\
Right slepton mass matrix, imaginary part:\\
 0.00000E+00  \> 0.00000E+00 \>  0.00000E+00\\
 0.00000E+00 \>  0.00000E+00 \>  0.00000E+00\\
 0.00000E+00 \> 0.00000E+00  \> 0.00000E+00\\
\\
Slepton LR mixing matrix, real part:\\
-9.00010E-03  \> 0.00000E+00  \> 0.00000E+00\\
 0.00000E+00  \> -1.86094E+00  \> 0.00000E+00\\
 0.00000E+00  \> 0.00000E+00 \>  -3.12978E+01\\
Slepton LR mixing matrix, imaginary part:\\
 0.00000E+00  \> 0.00000E+00 \>  0.00000E+00\\
 0.00000E+00 \>  0.00000E+00  \> 0.00000E+00\\
 0.00000E+00 \>  0.00000E+00 \>  0.00000E+00\\
\\
Left squark mass matrix, real part:\\
 2.50003E+05  \> 0.00000E+00 \>  0.00000E+00\\
 0.00000E+00  \> 2.50005E+05  \> 4.00010E+03\\
 0.00000E+00 \>  4.00010E+03 \>  1.60005E+05\\
Left squark mass matrix, imaginary part:\\
 0.00000E+00  \> 0.00000E+00  \> 0.00000E+00\\
 0.00000E+00 \>  0.00000E+00 \> -2.00005E+03\\
 0.00000E+00 \>  2.00005E+03 \>  0.00000E+00\\
\\
Right up-squark mass matrix, real part:\\
 2.02498E+05  \> 0.00000E+00 \>  0.00000E+00\\
 0.00000E+00 \>  2.02496E+05  \> 0.00000E+00\\
 0.00000E+00 \>  0.00000E+00 \>  3.99988E+04\\
Right up-squark mass matrix, imaginary part:\\
 0.00000E+00  \> 0.00000E+00  \> 0.00000E+00\\
 0.00000E+00 \>  0.00000E+00 \>  0.00000E+00\\
 0.00000E+00 \>  0.00000E+00 \>  0.00000E+00\\
\\
Right down-squark mass matrix, real part:\\
 3.02497E+05 \>  0.00000E+00 \>  0.00000E+00\\
 0.00000E+00 \>  3.02494E+05  \> 0.00000E+00\\
 0.00000E+00 \>  0.00000E+00  \> 8.99973E+04\\
Right down-squark mass matrix, imaginary part:\\
 0.00000E+00  \> 0.00000E+00  \> 0.00000E+00\\
 0.00000E+00 \>  0.00000E+00 \>  0.00000E+00\\
 0.00000E+00  \> 0.00000E+00 \>  0.00000E+00\\
\\
Up-squark LR mixing matrix, real part:\\
 6.18168E-03 \>  0.00000E+00 \> 0.00000E+00\\
 0.00000E+00 \> 1.79332E+00 \> 0.00000E+00\\
 0.00000E+00\>  0.00000E+00 \> 2.71002E+02\\
Up-squark LR mixing matrix, imaginary part:\\
 0.00000E+00 \> 0.00000E+00 \> 0.00000E+00\\
 0.00000E+00 \> 0.00000E+00\>  0.00000E+00\\
 0.00000E+00 \> 0.00000E+00 \> 0.00000E+00\\
\\
Down-squark LR mixing matrix, real part:\\
-1.19597E-01 \>  0.00000E+00 \> 0.00000E+00\\
 0.00000E+00 \> -1.87938E+00 \>  0.00000E+00\\
 0.00000E+00 \> 0.00000E+00 \> -5.56628E+01\\
Down-squark LR mixing matrix, imaginary part:\\
 0.00000E+00 \>  0.00000E+00 \> 0.00000E+00\\
 0.00000E+00 \> 0.00000E+00 \>  0.00000E+00\\
 0.00000E+00 \>  0.00000E+00 \>  0.00000E+00\\
~~~~~~~~~~~~~~~~~~~~~~~~~~~~~~~~~~~~~~~\=~~~~~~~~~~~\=~~~~~~~~~~~\=~~~~~~~~~~~\= \\
******* Particle masses in GeV: *******\\
** Fermion masses **\\
Charged lepton masses         \>        5.110E-04 \> 1.057E-01\>  1.777E+00\\
Running u quark masses at $m_t$ scale \> 2.220E-03 \> 6.440E-01\>  1.632E+02\\
Running d quark masses at $m_t$ scale\>  3.885E-03\>  6.104E-02 \> 2.737E+00\\
\\
** Higgs masses **\\
Tree level (H,h,A,H+):  \>  2.010E+02 \> 8.893E+01 \> 2.000E+02 \> 2.156E+02\\
1-loop, EPA approximation (H,h):\> 2.005E+02 \> 1.137E+02\\
\\
** Tree level SUSY masses **\\
~~~~~~~~~~~~~~~~~~~~~\=~~~~~~~~~~~\=~~~~~~~~~~~\=~~~~~~~~~~~\=~~~~~~~~~~~\= ~~~~~~~~~~~\=\\
Sneutrino masses  \>  2.897E+02 \> 2.931E+02\>  2.965E+02\\
Slepton masses   \>   2.953E+02 \> 3.028E+02\>  3.030E+02 \> 3.037E+02 \> 3.038E+02 \> 3.114E+02\\
U squark masses  \>   2.178E+02\>  4.486E+02 \> 4.487E+02 \> 4.489E+02\>  4.971E+02\>  4.974E+02\\
D squark masses   \>  2.999E+02 \> 4.049E+02\>  5.035E+02\>  5.037E+02 \> 5.505E+02\>  5.505E+02\\
Chargino masses  \>   1.552E+02\>  2.808E+02\\
Neutralino masses \>  8.865E+01\>  1.584E+02 \> 2.322E+02 \> 2.808E+02\\
Gluino mass     \>    6.000E+02
\end{tabbing}
}
\end{small}

The output for the physical observables is printed on the standard
output, usually the computer screen of the console.  It should look
like:
\begin{small} 
{\tt 
\begin{tabbing}
Electric dipole moments:~\=~~~~~~~~~~~~~~~~\=\\
Electron EDM \> =  4.7256E-25\\
Muon EDM    \>  =  9.7726E-23\\
Tau EDM     \>  =  1.6425E-21\\
Neutron EDM  \> =  5.9331E-24\\
\\
Neutrino K decays:\\
BR(K\_L$^0$ -> pi$^0$ vv) \> =  2.8555E-11\\
BR(K$^+$   -> pi$^+$ vv) \> =  7.3932E-11\\
\\
Leptonic B decays:\\
BR(B\_d -> mu$^+$ mu$^-$) \> =  1.2012E-10\\
BR(B\_s -> mu$^+$ mu$^-$) \> =  4.7395E-09\\
\\
B -> X\_s photon decay:\\
BR(B -> X\_s gamma) \> =  2.5756E-04\\
\\
KK mixing:\\
eps\_K \> =  2.3366E-03\\
Delta m\_K \> =  2.4362E-15\\
\\
DD mixing:\\
Delta m\_D \> =  1.6656E-17\\
\\
BB mixing:\\
Delta m\_B\_d \> =  3.6999E-13\\
Delta m\_B\_s \> =  1.3242E-11
\end{tabbing}
}
\end{small}

\end{appendix}

\newpage

\noindent{\bf PROGRAM SUMMARY}\\
\begin{small}
{\em Manuscript Title:}~ \code: a computational tool for FCNC and CP-violating processes in the MSSM\\
{\em Authors:}~ J.~Rosiek, P.~H.~Chankowski, A.~Dedes, S. J\"ager, P.~Tanedo\\
{\em Program Title:}~ \code \\
{\em Journal Reference:}                                      \\
%
{\em Catalogue identifier:}                                   \\
%
{\em Licensing provisions:}~ None\\
%
{\em Programming language:}~Fortran 77\\
{\em Operating system:}~Any, tested on Linux\\
%
{\em Keywords:}~Supersymmetry, $K$ physics, $B$ physics, rare decays, 
CP-violation   \\
{\em PACS:}~12.60.Jv, 13.20.He\\
{\em Classification:}~11.6 Phenomenological and Empirical Models and Theories\\
{\em Nature of problem:}\\
Predicting CP-violating observables, meson mixing parameters and
branching ratios for set of rare processes in the general R-parity
conserving MSSM.   \\
{\em Solution method:}\\ 
We use standard quantum theoretical methods to calculate Wilson
coefficients in MSSM and at one loop including QCD corrections at
higher orders when this is necessary and possible.  The input
parameters can be read from an external file in SLHA format.\\
{\em Restrictions:}\\
The results apply only to the case of MSSM with R-parity
conservation.\\
{\em Unusual features:}\\
{\em Running time:}\\
For single parameter set approximately 1s in {\tt double precision} on
a PowerBook Mac G4\\
{\em References:}

\begin{refnummer}

\item M.~Misiak, S.~Pokorski and J.~Rosiek, {\sl ``Supersymmetry and
  the FCNC effects''} Adv.\ Ser.\ Direct.\ High Energy Phys.\ {\bf 15}
  (1998) 795 [arXiv:hep-ph/9703442].

\item S.~Pokorski, J.~Rosiek and C.~A.~Savoy,
  Nucl.\ Phys.\ B {\bf 570} (2000) 81 [arXiv:hep-ph/9906206].

\item A.~J.~Buras, P.~H.~Chankowski, J.~Rosiek and L.~Slawianowska,
  Nucl.\ Phys.\ B {\bf 619} (2001) 434 [arXiv:hep-ph/0107048].

\item A.~J.~Buras, P.~H.~Chankowski, J.~Rosiek and L.~Slawianowska,
  Nucl.\ Phys.\ B {\bf 659} (2003) 3 [arXiv:hep-ph/0210145].

\item A.~J.~Buras, T.~Ewerth, S.~Jager and J.~Rosiek,
  Nucl.\ Phys.\ B {\bf 714} (2005) 103 [arXiv:hep-ph/0408142].

\item A.~Dedes, J.~Rosiek and P.~Tanedo,
  Phys.\ Rev.\ D {\bf 79} (2009) 055006 [arXiv:0812.4320 [hep-ph]].

\end{refnummer}

\end{small}

\end{document}